\documentclass[sn-mathphys-num]{sn-jnl}

\usepackage{graphicx}%
\usepackage{multirow}%
\usepackage{amsmath,amssymb,amsfonts}%
\usepackage{amsthm}%
\usepackage{mathrsfs}%
\usepackage[title]{appendix}%
\usepackage{xcolor}%
\usepackage{textcomp}%
\usepackage{manyfoot}%
\usepackage{booktabs}%
\usepackage{algorithm}%
\usepackage{algorithmicx}%
\usepackage{algpseudocode}%
\usepackage{listings}%

\theoremstyle{thmstyleone}%
\theoremstyle{thmstyletwo}%

\theoremstyle{thmstylethree}%

\begin{document}

\title[Article Title]{Study of the tracking efficiency of charged pions at BESIII}

\author*[1]{\fnm{Fang Liu}}\email{liufang@ihep.ac.cn}

\author[1,3]{\fnm{Xiao-Bin Ji}}\email{jixb@ihep.ac.cn}
\equalcont{These authors contributed equally to this work.}

\author[1,3]{\fnm{Sheng-Sen Sun}}\email{sunss@ihep.ac.cn}
\equalcont{These authors contributed equally to this work.}

\author[1,3]{\fnm{Huai-Min Liu}}
\author[1,3]{\fnm{Shuang-Shi Fang}}
\author[2]{\fnm{Xiao-Ling Li}}
\author[1]{\fnm{Tong Chen}}
\author[1,3]{\fnm{Xin-Nan Wang}}
\author[1,3]{\fnm{Ming-Run Li}}
\author[1]{\fnm{Liang-Liang Wang}}
\author[1]{\fnm{Ling-Hui Wu}}
\author[1,3]{\fnm{Ye Yuan}}
\author[1]{\fnm{Yao Zhang}}
\author[1]{\fnm{Wen-Jing Zhu}}

\affil*[1]{\orgname{Institute of High Energy Physics}, 
\orgaddress{
\city{Beijing},\postcode{100049}, 
\country{China}}}

\affil[2]{\orgname{Shandong University}, \orgaddress{
\city{Jinan}, \postcode{250100}, \state{Shandong Province}, \country{China}}}

\affil[3]{\orgname{University of Chineses Academy of Sciences}, 
\orgaddress{
\city{Beijing},\postcode{100049}, 
\country{China}}}


\abstract{
Using $(10087 \pm 44) \times 10^6$ $J/\psi$ events collected with the BESIII detector in 2009, 2012, 2018 and 2019, the tracking efficiency of charged pions is studied using the decay $J/\psi \rightarrow \pi^+ \pi^- \pi^0$.
The systematic uncertainty of the tracking efficiency and the corresponding correction factors for charged pions are evaluated, in bins of transverse momentum and polar angle of the charged pions.}




\maketitle

\section{Introduction}\label{sec1}
The BESIII detector~\cite{Ablikim:2009aa} at the Beijing Electron-Positron Collider (BEPCII) ~\cite{Yu:IPAC2016-TUYA01}
has collected the largest sample of $J/\psi$ events at an $e^{+} e^{-}$ collider, leading to a significant reduction of the statistical uncertainties for many measurements. It is crucial to estimate the systematic uncertainties more rigorously to align with the unprecedented statistical precision. The systematic uncertainties are primarily due to differences between data and Monte Carlo (MC) simulation. Track reconstruction of charged particles is a key source of the systemic uncertainties for many measurements~\cite{PhysRevLett.129.131801,zhaozh2024,PhysRevD.103.072006}. Therefore the estimation of the systematic uncertainty related to tracking is a critical aspect at BESIII.

The key tracking sub-detector within the BESIII detector is a small-celled, helium-based main drift
chamber (MDC) with 43 layers of wires, which has a
geometrical acceptance of $93\%$ of $4 \pi$ and provides momentum
 and ionization energy loss ($dE/dx$) measurements for charged particles.
The other sub-detectors include  a plastic
scintillator time-of-flight (TOF) system, a CsI(Tl) electromagnetic calorimeter (EMC) and a muon counter
 (MUC) made of Resistive Plate Chambers.

In this paper, the tracking efficiency of charged pions is studied with a high purity pion sample, which is selected from the decay $J/\psi \rightarrow \pi^+ \pi^- \pi^0$
with
high statistics, broad range of  momentum  and angular distributions.
To better understand the tracking efficiency the dataset with $(10087 \pm 44) \times 10^6$ $J/\psi$ events~\cite{yanghx}
 is divided into four subsets due to the detector status and data taken condition, corresponding to the years of 2009, 2012, 2018, and 2019.
 To investigate the event selection criteria and determine the
 detection efficiency, the signal MC samples are
 generated with DIY generator 
 which is based on covariant tensor formalism~\cite{bszou2003}
 and modeled with the
 partial wave analysis amplitudes of the decay
 $J/\psi \rightarrow \pi^+ \pi^- \pi^0$. Here the excited states such
 as $\rho(1450)$ in the $\pi\pi$ invariant mass and the interferences
 among three decay modes
 $J/\psi \rightarrow \rho^{\pm(0)}\pi^{\mp(0)}$ and phase space are
 considered, and four MC samples are simulated to match four data
 subsets, separately.  To investigate potential background, an
 inclusive MC sample of $1.0 \times 10^{10} J/\psi$ events is
 analyzed, which includes the production of the $J/\psi$ resonance and
 the continuum processes. 
The simulation models the beam-energy 
 spread and initial-state radiation (ISR)  in  $e^{+}e^{-}$ annihilations with the 
 generator \textsc{kkmc}~\cite{kkmc2000,kkmc2001}.
 The known
 particle decays are modeled with
 \textsc{evtgen}~\cite{2001Lange,2008ping} using branching fractions
 taken from the Particle Data Group~\cite{2022pdg}, while the
 remaining unknown $J/\psi$ decays are estimated with
 \textsc{lundcharm}~\cite{2000chen,lund2014}. The interaction of the generated particles with the
detector, and its response, are implemented using the Geant4 toolkit ~\cite{geant4,geant4-2003} as described
in Ref. ~\cite{Ablikim:2009aa}.
 To better model data,
 the signal events in the inclusive MC sample are replaced with the
 signal MC events of $J/\psi \rightarrow \pi^+ \pi^- \pi^0$.

 In this paper, the tracking efficiency of charged pion is studied
 with respect to transverse momentum $p_t$, polar angle $\theta$ and
 azimuthal angle $\phi$ in the MDC. The tracking efficiency is found
 to be sensitive to $p_t$ and $\cos\theta$ due to different tracking bendings and hit positions, but insensitive to $\phi$.
 Two-dimensional tracking efficiencies as a function of $p_t$ and
 $\cos \theta$ are evaluated for both data and MC samples. To correct
 the MC efficiencies to those of the data, the corresponding
 two-dimensional correction factors and their uncertainties are
 derived. These results are important to reduce the systematic
 uncertainty of tracking for charged pions.

\section{Definition of tracking efficiency}
\indent The tracking efficiency of pion is obtained by the tag-and-probe method using the control sample of $J/\psi \rightarrow \pi^+ \pi^- \pi^0$.  The tag charged pion is required to satisfy stringent identification criteria and the probe charged pion is treated as a missing track which is only determined by the recoiling information in the control sample selection. Then in the selected control sample, one can account for the number of reconstructed probe pions.

The tracking efficiency ($\epsilon$) for charged pions is defined as
\begin{equation}
\label{eqcount_eff}
  \epsilon =\frac{N}{N^{\prime}},
\end{equation}
where
$N^{\prime}$ represents the number of events with one charged track or
two charged tracks with zero total net charge, and $N$ represents the
number of events with the missing track found successfully with its
charge property matched. Both $N$ and $N^{\prime}$ are obtained by
counting.
The uncertainty of the tracking efficiency ($\sigma_{\epsilon}$) is determined by
 \begin{equation}
\sigma_{\epsilon}=\sqrt{\frac{\epsilon(1-\epsilon)}{N^{\prime}}}.
\end{equation}
The relative difference of tracking efficiencies between MC simulations
($\epsilon_{\rm MC}$) and data ($\epsilon_{\rm data}$) is defined as
\begin{equation}
\Delta\epsilon=\frac{\epsilon_{\rm MC}-\epsilon_{\rm data}}{\epsilon_{\rm MC}}.
\end{equation}
The correction factor of tracking efficiency for data over MC simulations is taken as
$1-\Delta\epsilon$. Then, the error of the correction factor is
determined as
\begin{equation}
\sigma_{\Delta\epsilon} = \frac{\epsilon_{\rm data}}{\epsilon_{\rm MC}} \times
\sqrt{(\frac{\sigma_{\epsilon_{\rm MC}}}{\epsilon_{\rm MC}})^{2}+(\frac{\sigma_{\epsilon_{\rm data}}}{\epsilon_{\rm data}})^{2}}.
\end{equation}

The candidate events are required to have at least one charged track and
only two photon candidates. The charged tracks are  
reconstructed from hits in the MDC within the polar angle range $|\cos\theta|<0.93$.
Their distance of closest approach to the interaction point (IP) must be
less than 10\,cm along the beam direction and less than 1\,cm in the
transverse plane.
Particle identification~(PID) for the tag charged track combines measurements
of the energy deposited in the MDC~(d$E$/d$x$) and the flight time in
the TOF to form likelihoods $\mathcal{L}(h)~(h=p,K,\pi)$ for each
hadron $h$ hypothesis.Tracks are identified as pions when the pion
hypothesis has the greatest likelihood
($\mathcal{L}(\pi)>\mathcal{L}(p)$ and
$\mathcal{L}(\pi)>\mathcal{L}(K)$).  Photon candidate is reconstructed
by isolated showers in the EMC and its energy is required to be
greater than 25 MeV in barrel ($|\cos\theta|<0.80$) and 50 MeV in end
cap ($0.86<|\cos\theta|<0.92$).  To exclude showers that originate
from charged tracks, the angle subtended by the EMC shower and the
position of the closest charged track at the EMC must be greater than
10 degrees as measured from the IP. To suppress electronic noise and
showers unrelated to the event, the difference between the EMC time
and the event start time is required to be within 700 ns.

\begin{figure}[h]
\centering
 \includegraphics[width=0.35\textwidth]{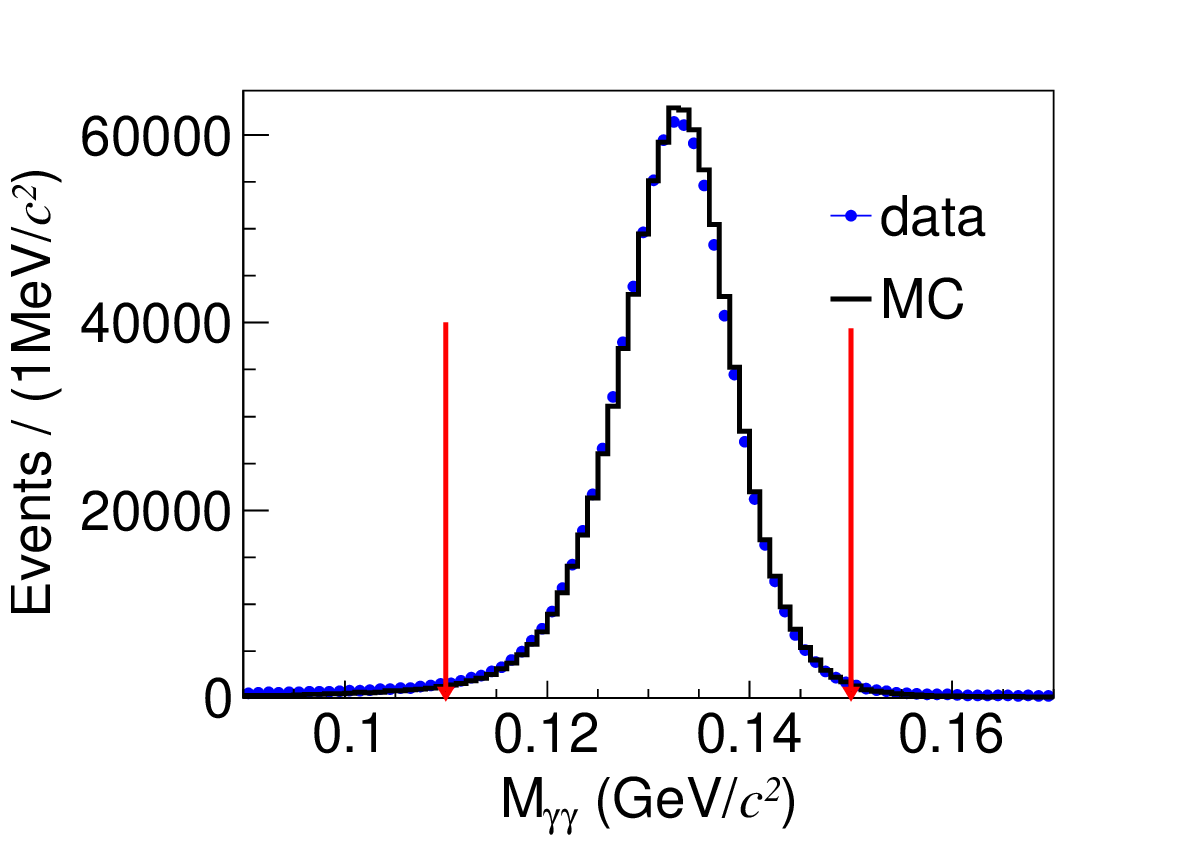}
  \includegraphics[width=0.35\textwidth]{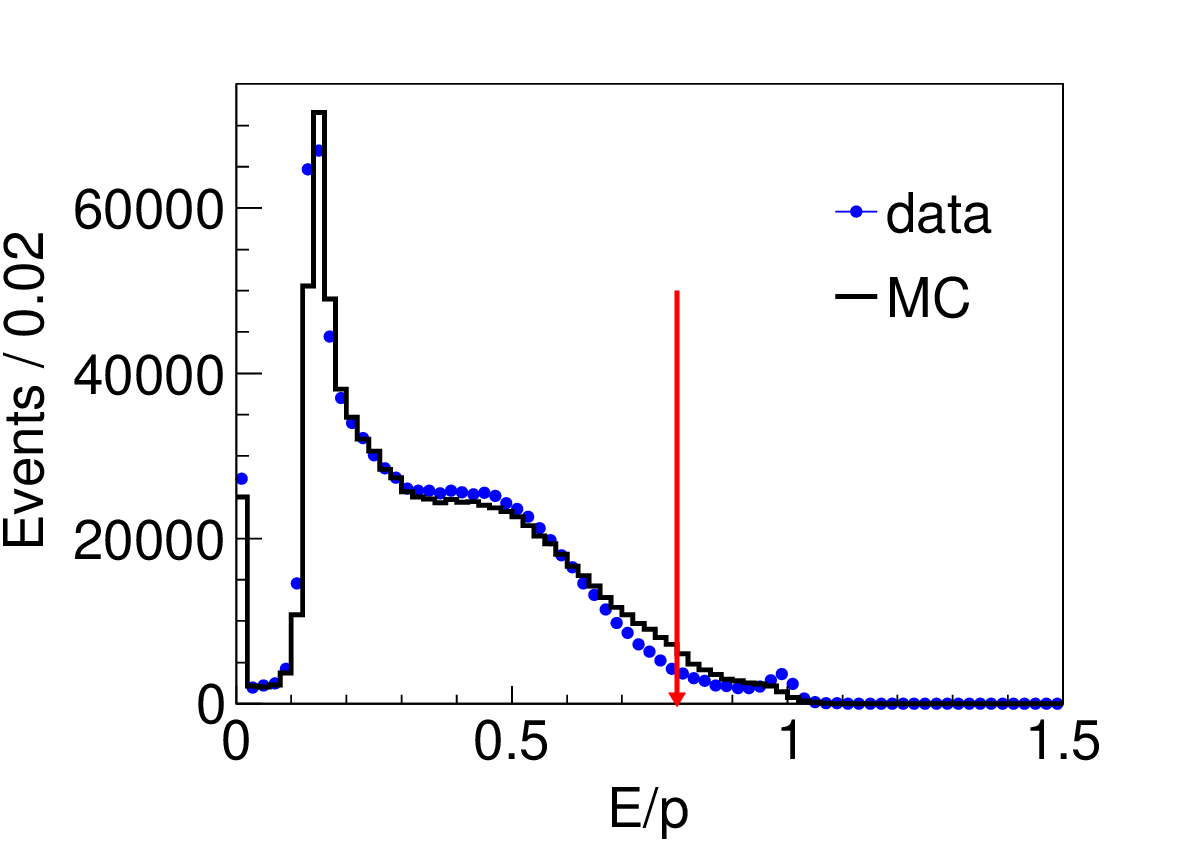}
    \put(-220,70){\bf (a)}  \put(-75,70){\bf (b)} \\
  \includegraphics[width=0.35\textwidth]{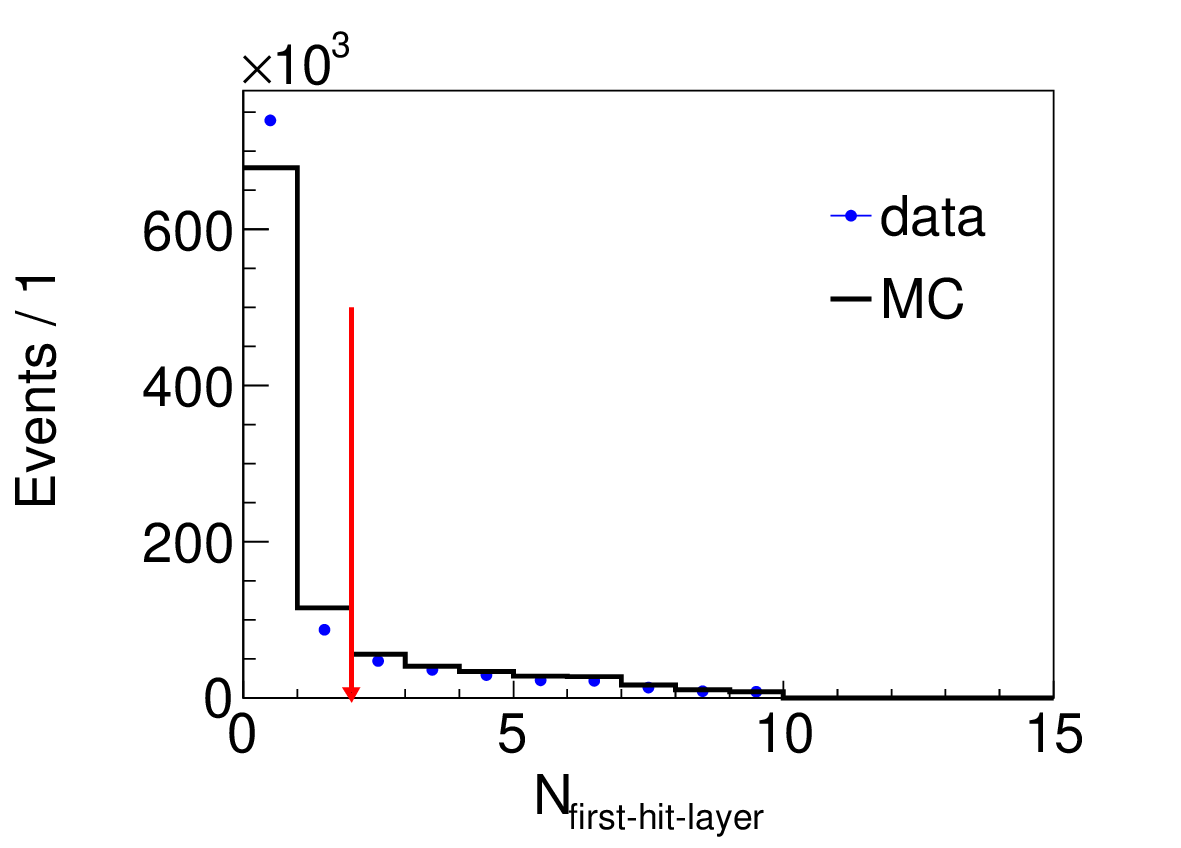}
  \includegraphics[width=0.35\textwidth]{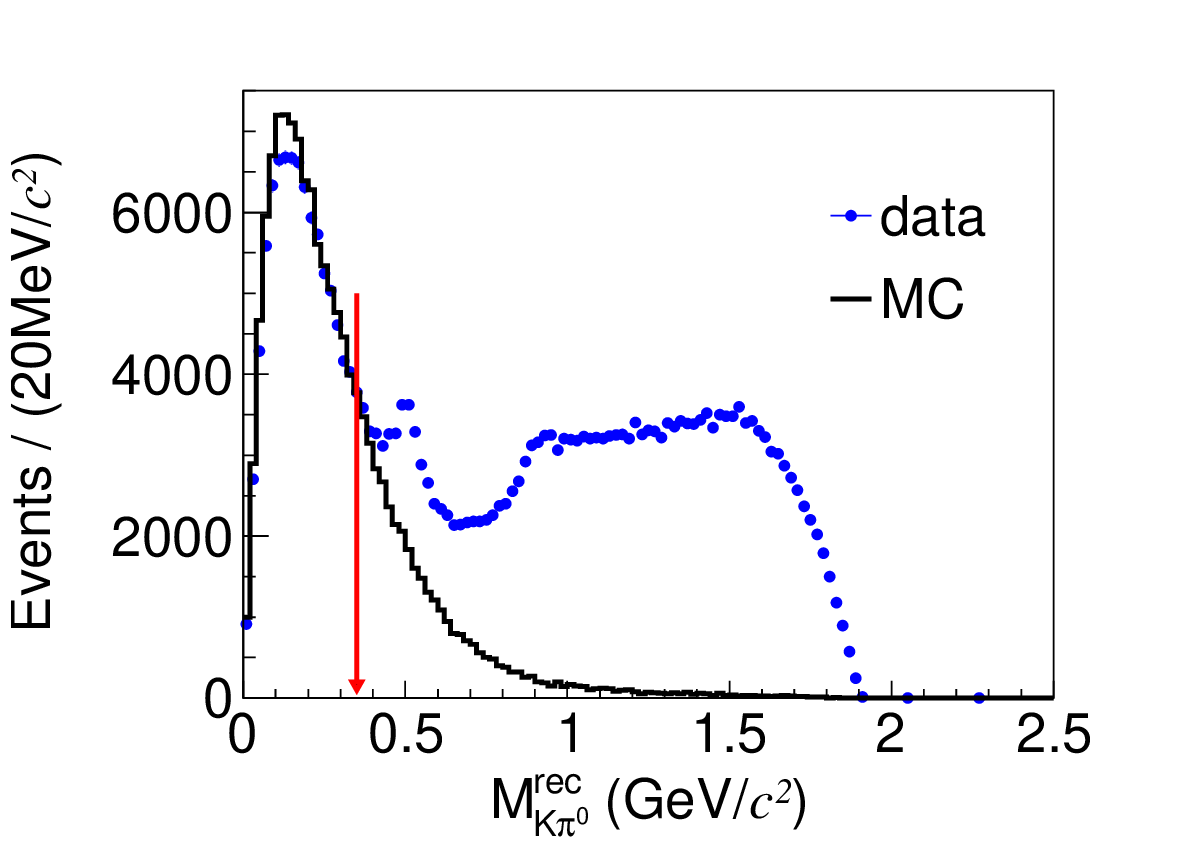}
   \put(-220,70){\bf (c)}  \put(-75,70){\bf (d)} \\
 \caption{The distributions of  (a) $M_{\gamma\gamma}$, (b) $E/p$, (c) $N_{\rm first-hit-layer}$ and (d) $M^{\rm rec}_{K\pi^0}$ from the selected data and MC samples.}
 \label{fig:track_pi0_decay_angle}
\end{figure}

\indent Candidate $\pi^0$ mesons are reconstructed from pairs of
photons with an invariant mass in the range of $0.110$
GeV$/c^{2}<M_{\gamma \gamma}<0.15$ GeV$/c^{2}$, as shown in
Fig.~\ref{fig:track_pi0_decay_angle}(a). The decay angle
$\cos\theta_{\gamma_1
  \gamma_2}=\frac{|E_{\gamma_1}-E_{\gamma_2}|}{p_{\pi^{0}}}$,
where $p_{\pi^{0}}$ is the momentum for the mother particle of two
photons, is required to be less than 0.95. The tag charged track is required to satisfy these additional identification criteria.
To suppress backgrounds from Bhabha events and $e^+ e^- \rightarrow \mu^+ \mu^-$, the ratio of the energy deposited in the EMC over its
momentum in MDC ($E/p$) and the first hit layer in the MUC ($N_{\rm first-hit-layer}$) associated to the tagged charged
track, as shown in Fig.~\ref{fig:track_pi0_decay_angle}(b) and Fig.~\ref{fig:track_pi0_decay_angle}(c), are
required to be less than 0.8 and 3, respectively.
If the tagged charged track is a kaon and mis-identified as a pion, the recoiling mass of the
charged track and two photons $M^{\rm rec}_{K\pi^0}$ in Fig.~\ref{fig:track_pi0_decay_angle}(d) is required to be less than $0.35$ GeV$/c^{2}$ to
remove the background events from $J/\psi \rightarrow K^+ K^- \pi^0$.

The previously mentioned selection criteria are suitable for $\pi^\pm$ samples, with the following selection being distinct. 
In the study of
$\pi^{+}$ tracking efficiency, the
$J/\psi\rightarrow \rho^{-} \pi^{+}$ mode is rejected with the
requirement $M_{\pi^{-}\pi^0}>1.0$ GeV$/c^{2}$ due to its relative
high background among these three decay modes $J/\psi \rightarrow \rho^{\pm(0)} \pi^{\mp(0)}$. Even so, the statistics is still high enough.
Similarly, the $J/\psi \rightarrow \rho^{+} \pi^{-}$ mode is not used
with the requirement of $M_{\pi^{+}\pi^0}>1.0$ GeV$/c^{2}$ in the
study of $\pi^{-}$ tracking efficiency.

\begin{figure}[h]
\centering
 \includegraphics[width=0.35\textwidth]{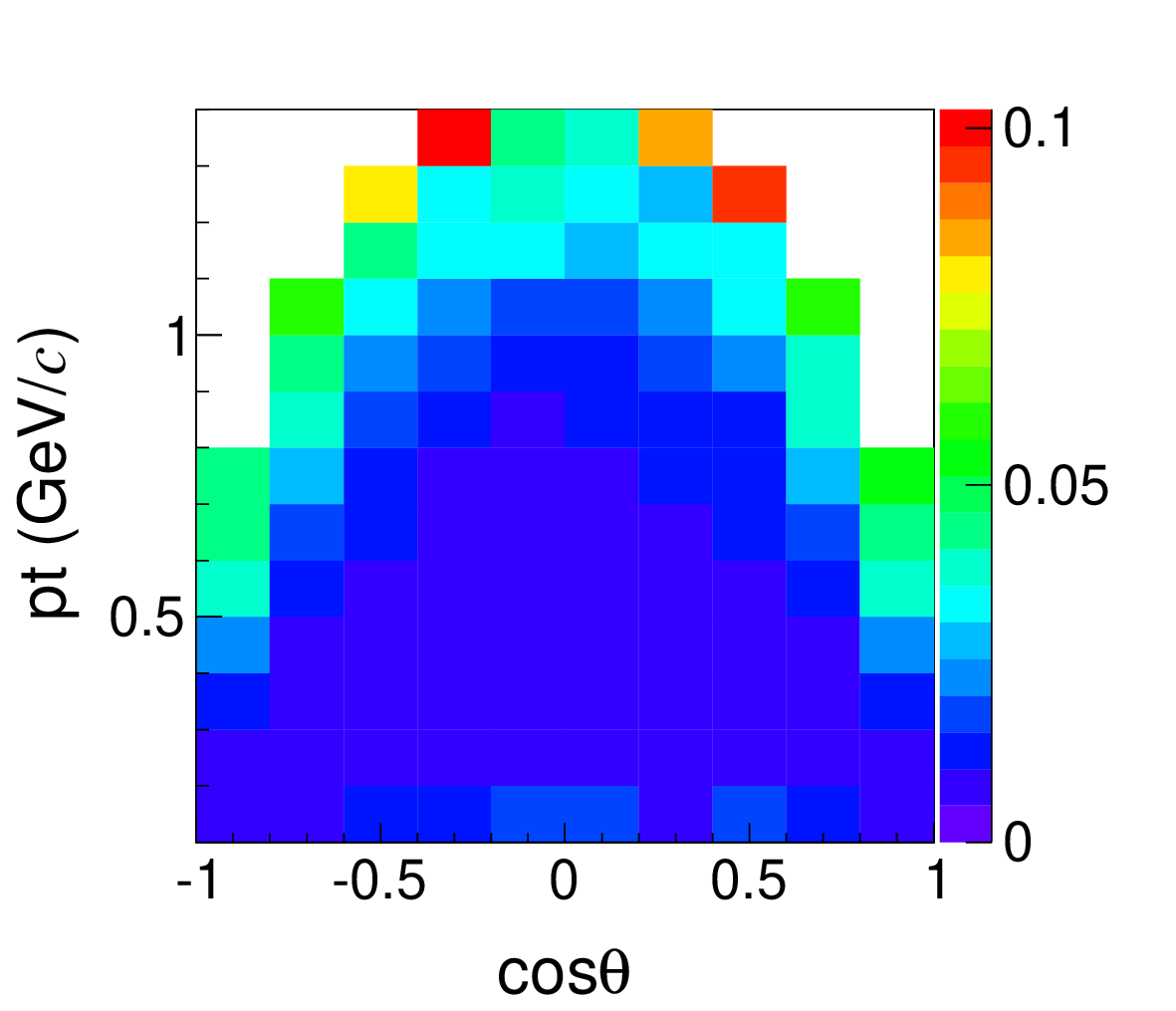}
 \caption{The distribution of background levels with respect to $p_t$ and $\cos\theta$ of the selected $\pi^+$ sample.}
 \label{fig:trk_bg_2d_level}
\end{figure}

The possible background events are estimated by analyzing the
inclusive MC sample of $1.0 \times 10^{10} J/\psi$ events. The main backgrounds come from $J/\psi$ decays into two charged
pions with the neutral tracks such as
$J/\psi \rightarrow \pi^+ \pi^- (\gamma/\gamma\pi^0/\pi^0\pi^0)$ and
neutron-antineutron pair $J/\psi \rightarrow \pi^+ \pi^- n\bar{n}$,
and the average background level is about $2.0\%$, as shown in
Fig.~\ref{fig:trk_bg_2d_level}. The non-pion background level is up to
about $0.4\%$ with slight difference in individual momentum ranges for
each $\cos\theta$ bin.

\section{Evaluation of Tracking Efficiency of Charged Pions}

\begin{figure*}[h]
\centering
 \includegraphics[width=0.32\textwidth]{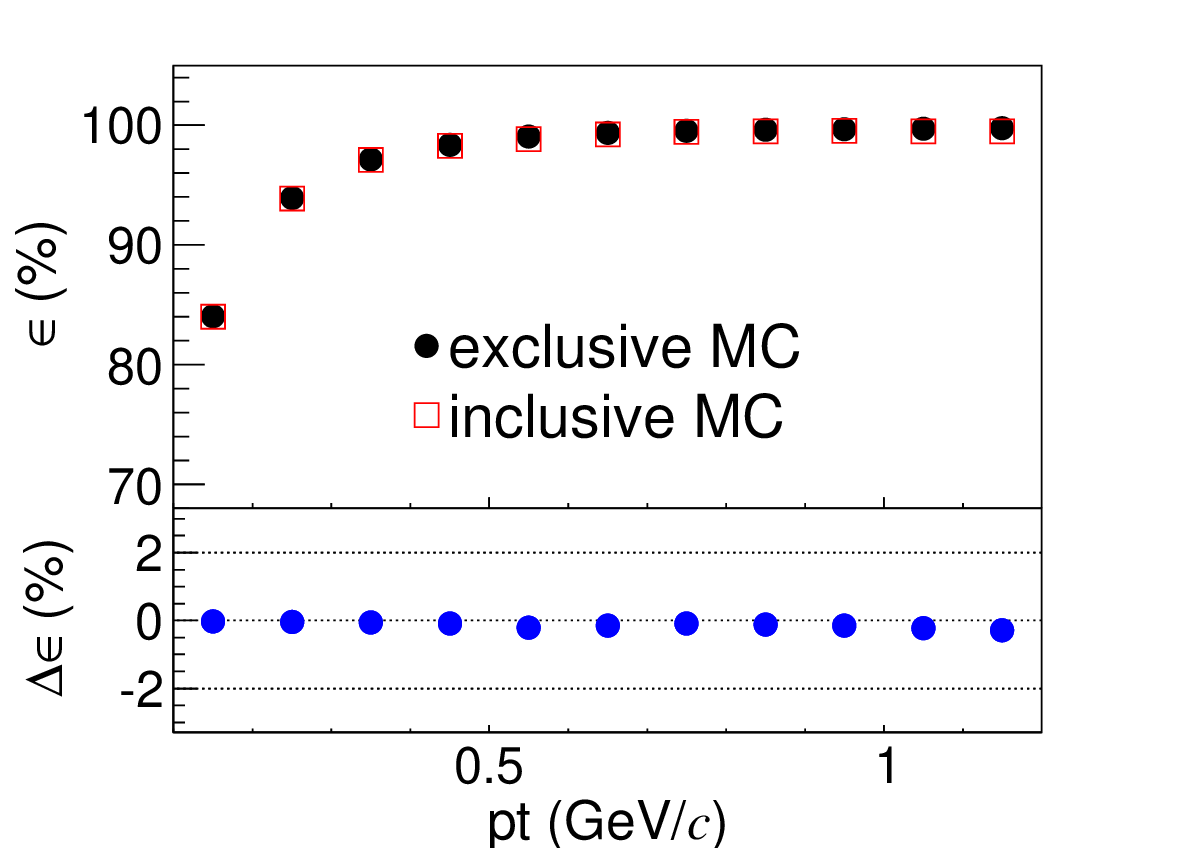}
  \includegraphics[width=0.32\textwidth]{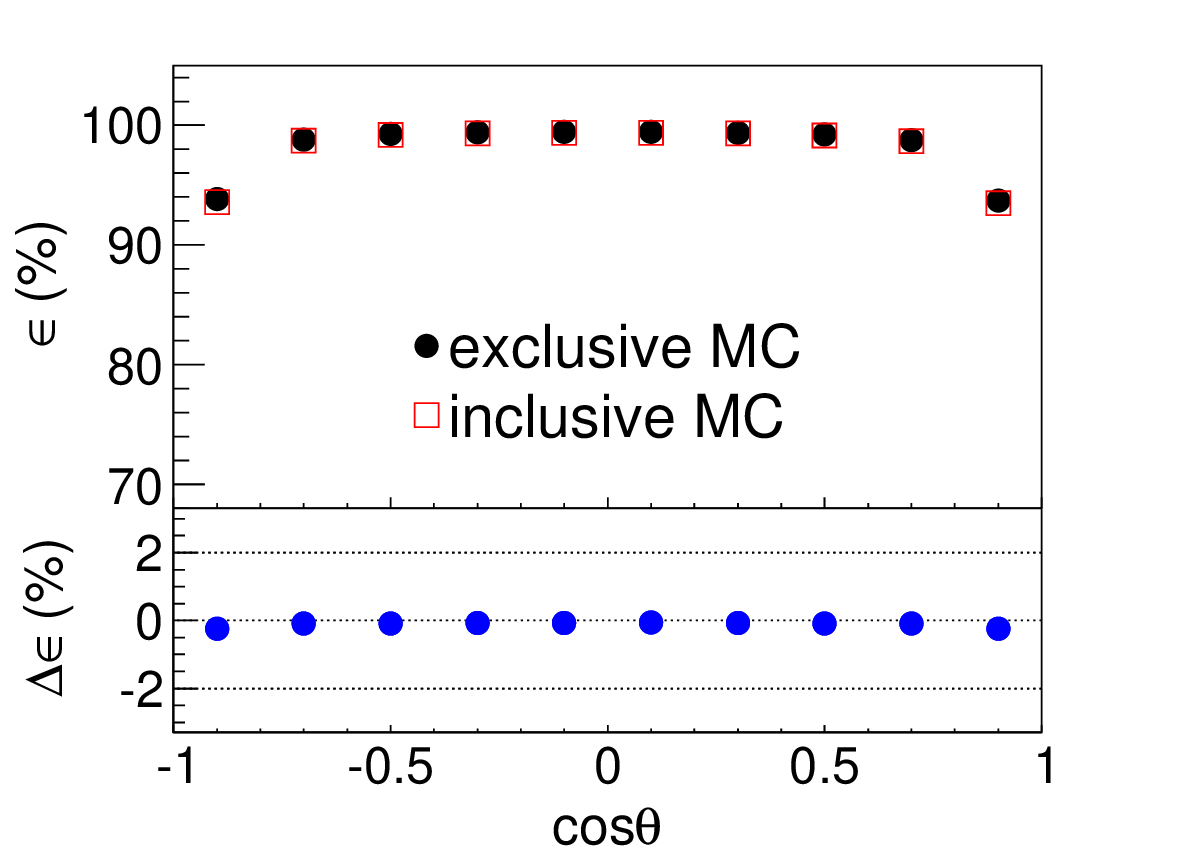}
  \includegraphics[width=0.32\textwidth]{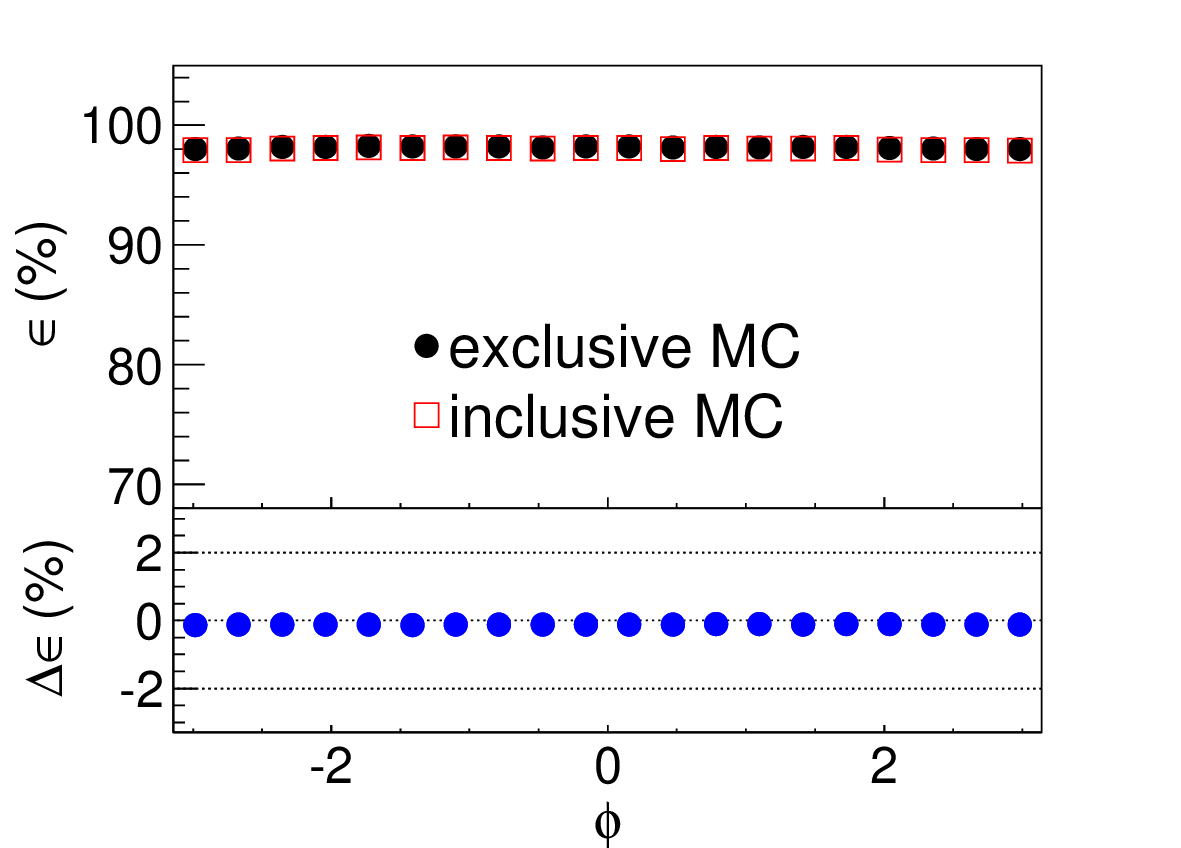}
 \put(-310,55){\bf (a)}\put(-195,55){\bf (b)}\put(-70,55){\bf (c)}\\
 \caption{ Pion tracking efficiencies versus (a) $p_t$, (b) $\cos\theta$ and (c) $\phi$ between inclusive MC and signal (exclusive) MC. Bottom panels show $\Delta \epsilon$,  defined as $\frac{\epsilon_{\rm inMC}-\epsilon_{\rm exMC}}{\epsilon_{\rm inMC}}$.}
 \label{fig:trk_pt_2mc_ex_in_2018}
\end{figure*}

The tracking efficiency of charged pions versus $p_t$, $\cos\theta$ and
$\phi$, for the inclusive MC and the
$J/\psi \rightarrow \pi^+ \pi^- \pi^0$ MC, are studied individually,
with results shown in Fig.~\ref{fig:trk_pt_2mc_ex_in_2018}. Small
difference between the two MC samples is visible in the end cap region
or the high momentum region from 1.1 GeV$/c$ to 1.2 GeV$/c$ due to background events. The inclusive MC sample including the background effect is used to study tracking efficiency and discussed later comparing real data.

\begin{figure*}[h]
\centering
 \includegraphics[width=0.32\textwidth]{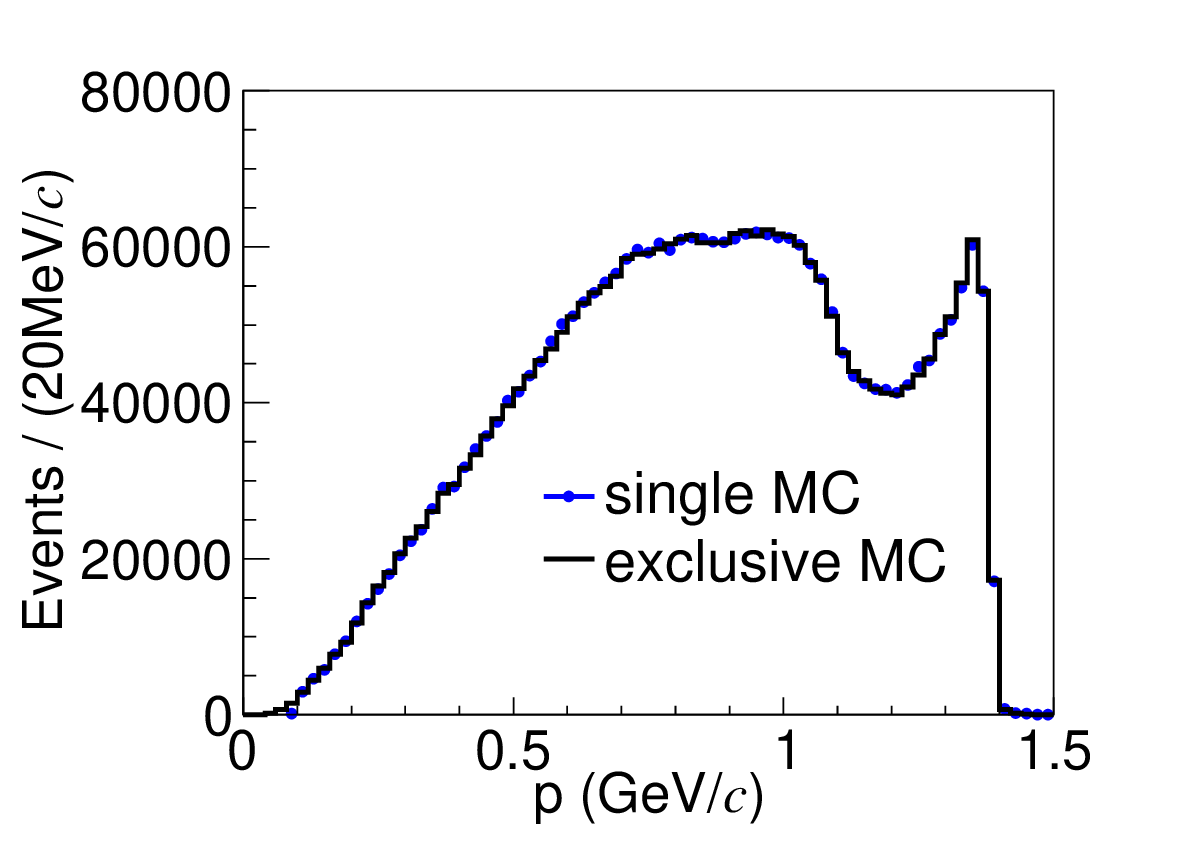}
 \includegraphics[width=0.32\textwidth]{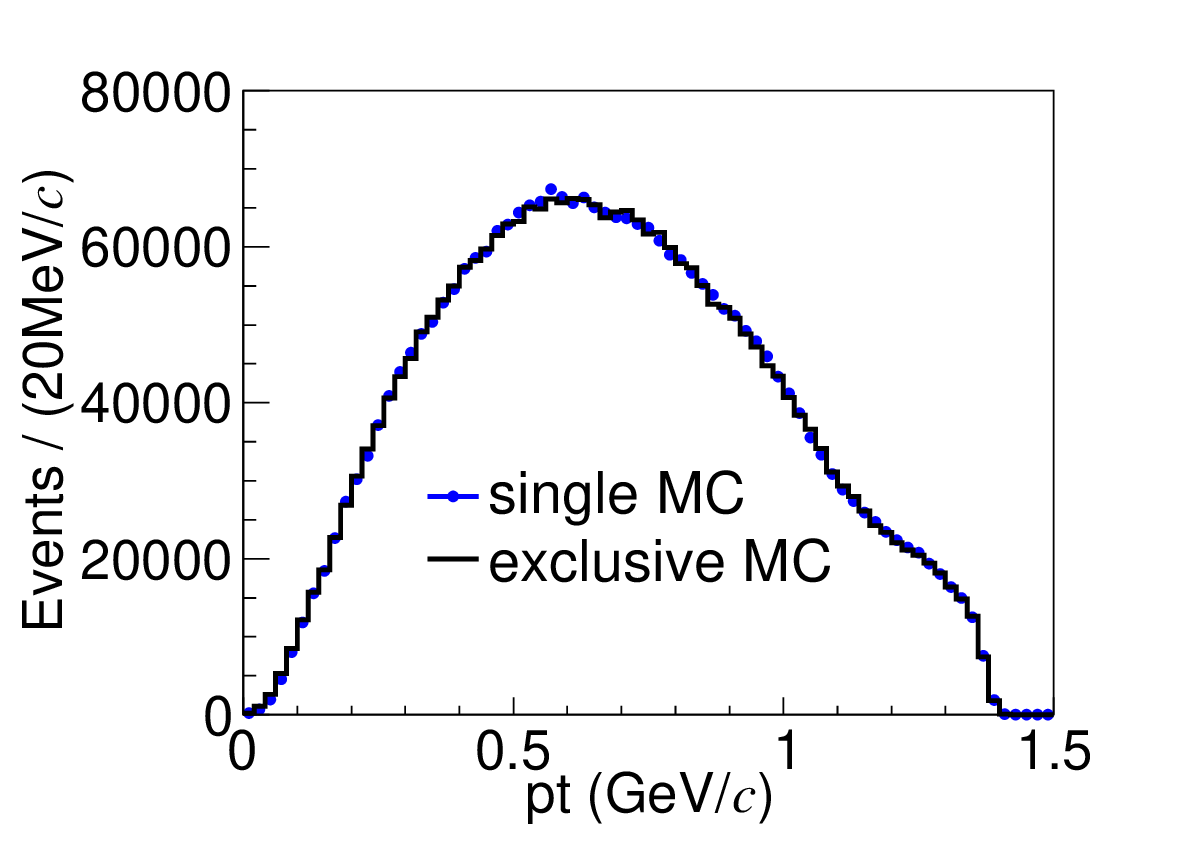}
 \includegraphics[width=0.32\textwidth]{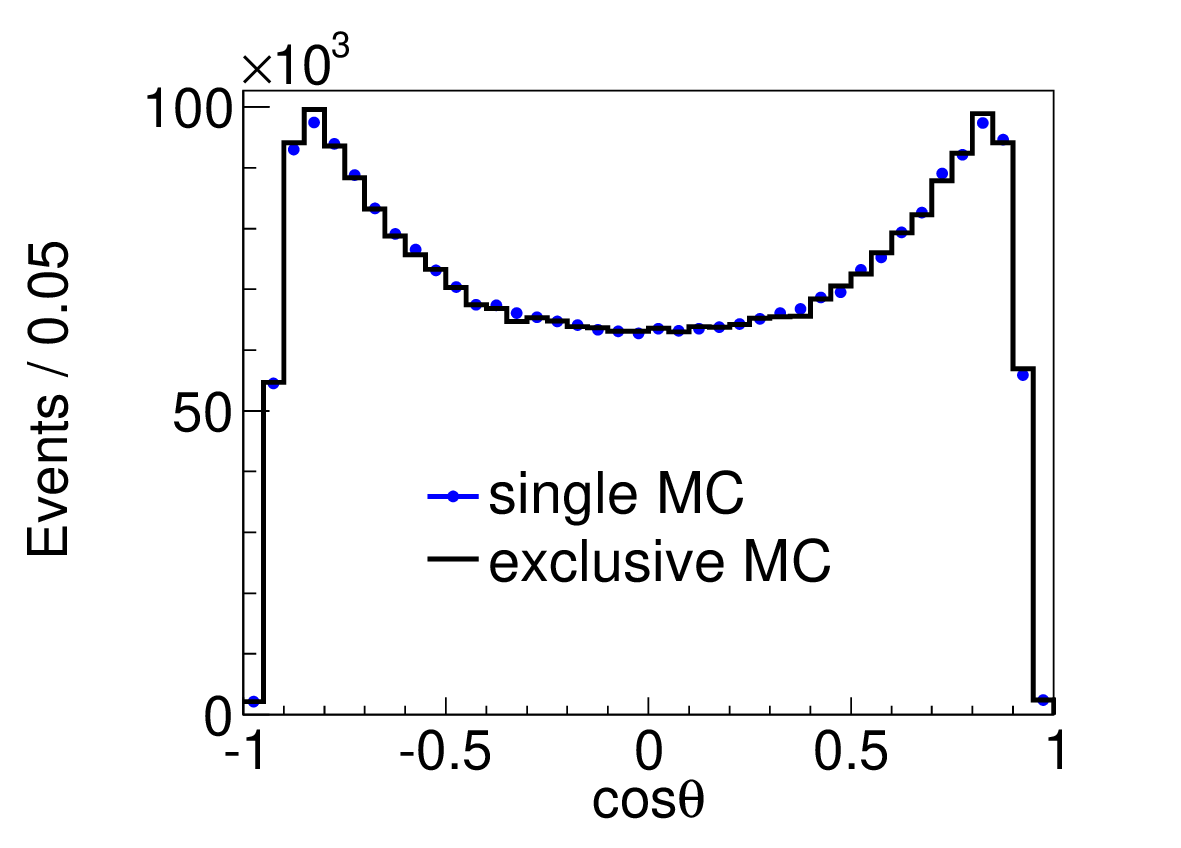}
  \put(-325,60){\bf (a)}\put(-210,60){\bf (b)}\put(-60,60){\bf (c)}\\
 \caption{ (a) Momentum, (b) $p_t$  and (c) $\cos\theta$  distributions of single pion track
 MC and exclusive MC.}
 \label{fig:momentum-single-exclusive-t0-mctruth-distribution}
\end{figure*}

\begin{figure*}[h]
\centering
 \includegraphics[width=0.32\textwidth]{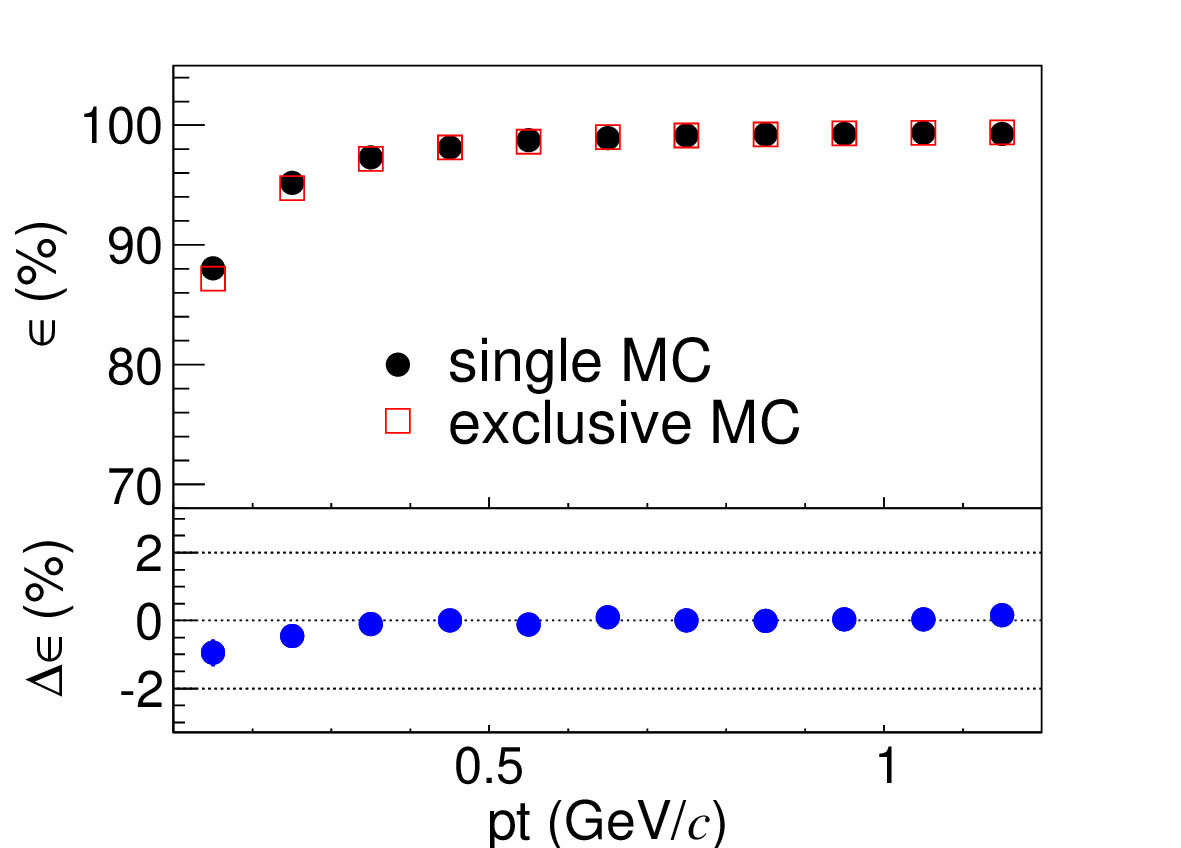}
  \includegraphics[width=0.32\textwidth]{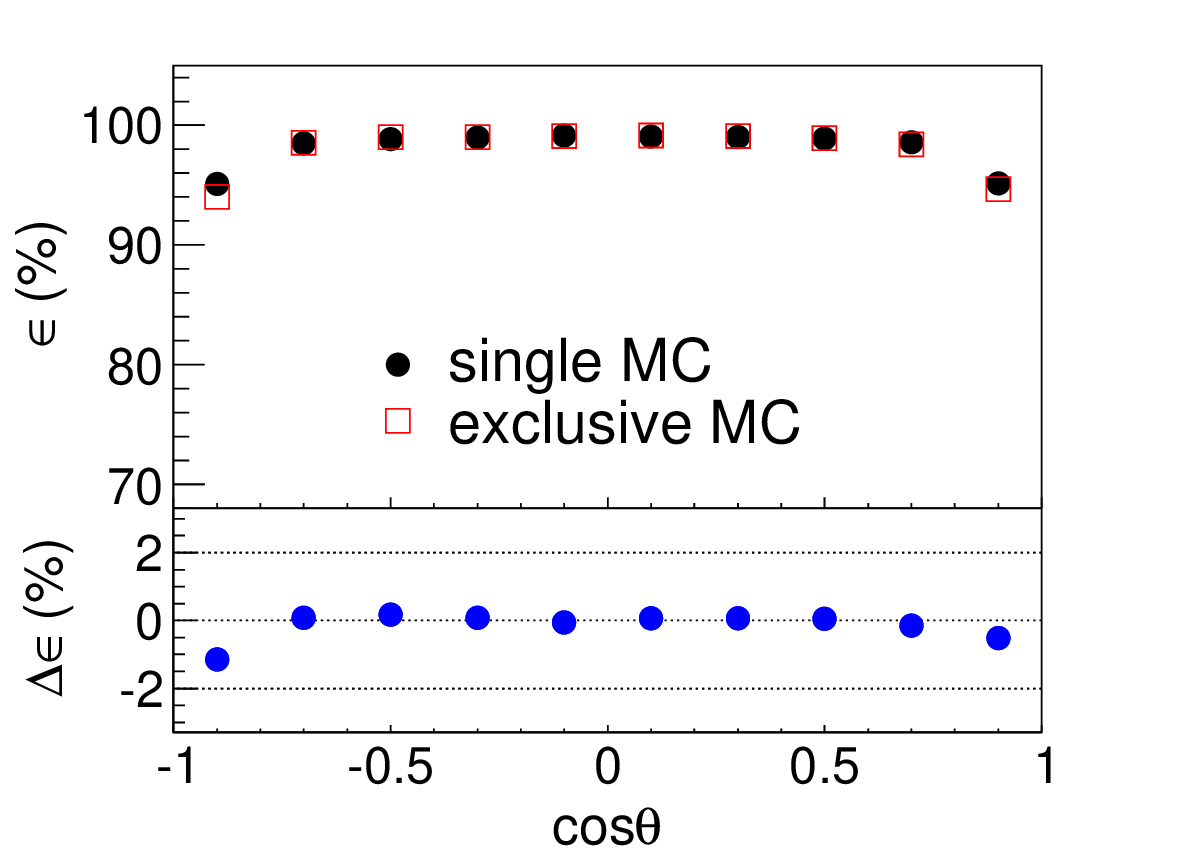}
 \put(-200,60){\bf (a)}\put(-75,60){\bf (b)}\\
 \caption{ Pion tracking efficiencies versus (a) $p_t$ and (b)
   $\cos\theta$ of single pion track MC and exclusive MC. Bottom
   panels show $\Delta \epsilon$, defined as
   $\frac{\epsilon_{\rm exMC}-\epsilon_{\rm single-MC}}{\epsilon_{\rm
       exMC}}$.}
 \label{fig:momentum-single-exclusive-t0-mctruth}
\end{figure*}

The tracking efficiencies are further validated by the tracking
efficiency from a single pion track MC sample, which is generated
uniformly in phase space and re-weighted according to the
distributions of the signal MC sample.  The momentum, transverse
momentum, and polar angle of these two MC samples are shown in
Fig.~\ref{fig:momentum-single-exclusive-t0-mctruth-distribution}, in
which good consistencies can be seen. The pion tracking efficiencies
versus $p_t$ and $\cos\theta$ for these two MC samples are shown in
Fig.~\ref{fig:momentum-single-exclusive-t0-mctruth}. They are almost
consistent with each other except a tiny difference in the low
transverse momentum and end cap regions.  Here the little difference
in low transverse momentum may be caused by the other charged
tracks. Therefore, no obvious bias is observed in event selection and
investigation.  For pion with low transverse momentum less than 0.6 GeV$/c$, the control
sample of $J/\psi \rightarrow p \bar{p} \pi^{+} \pi^{-}$ is 
preferable to $J/\psi \rightarrow \pi^+\pi^-\pi^0$ due to its smoother angular distribution and higher statistics. 

\begin{figure*}[h]
\centering
 \includegraphics[width=0.3\textwidth]{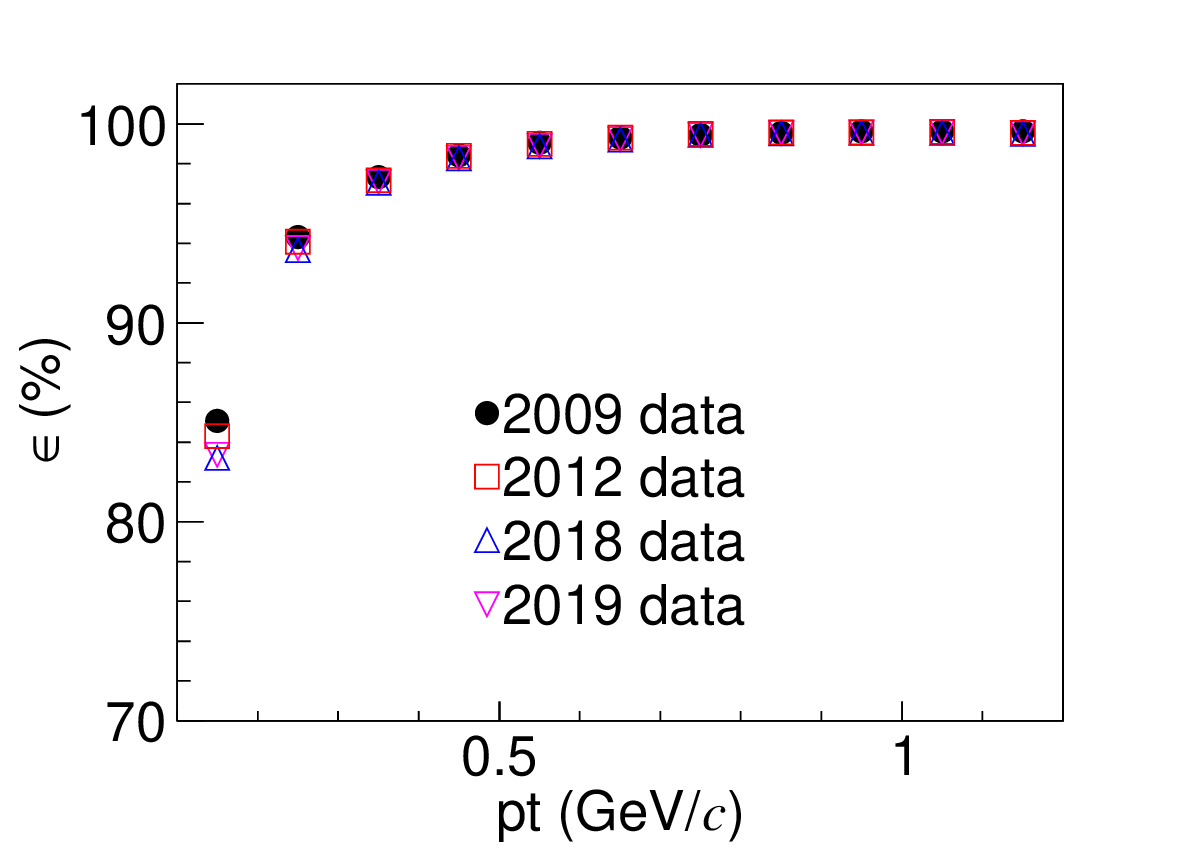}
\includegraphics[width=0.3\textwidth]{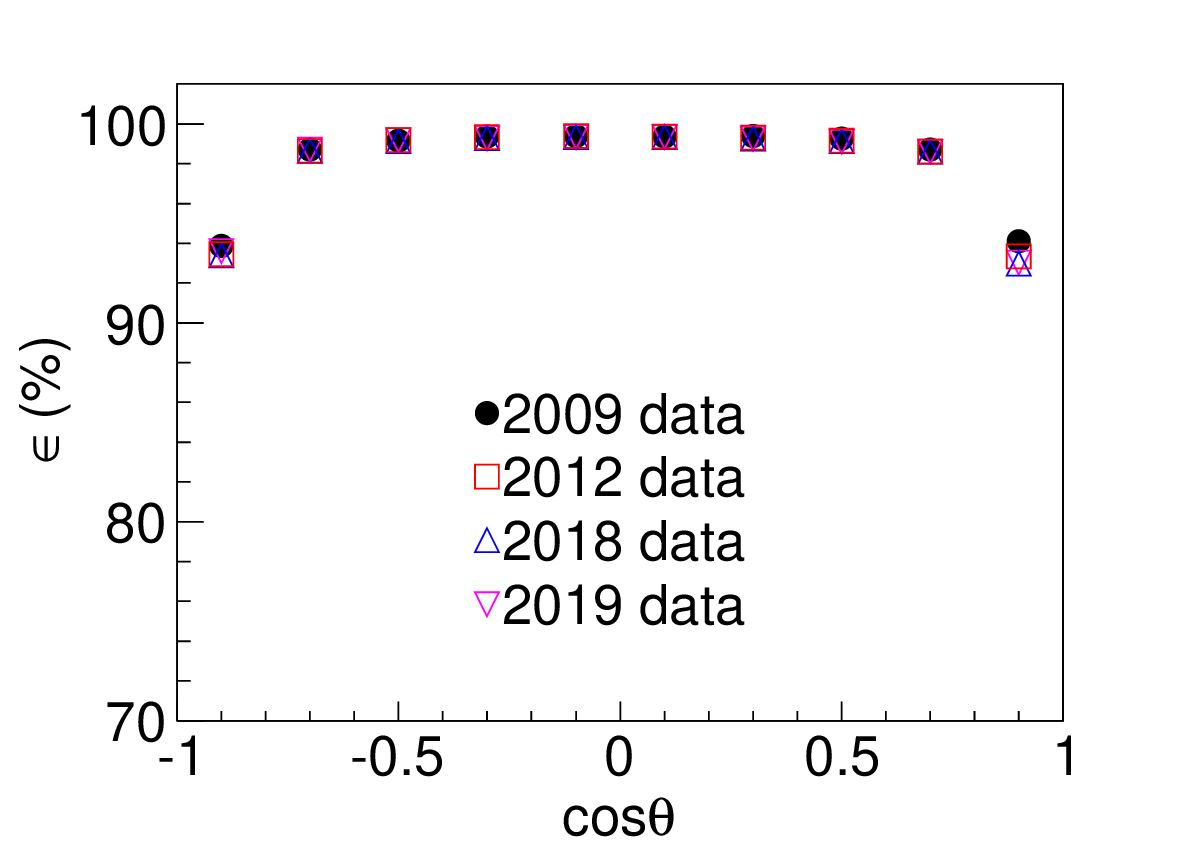}
\includegraphics[width=0.3\textwidth]{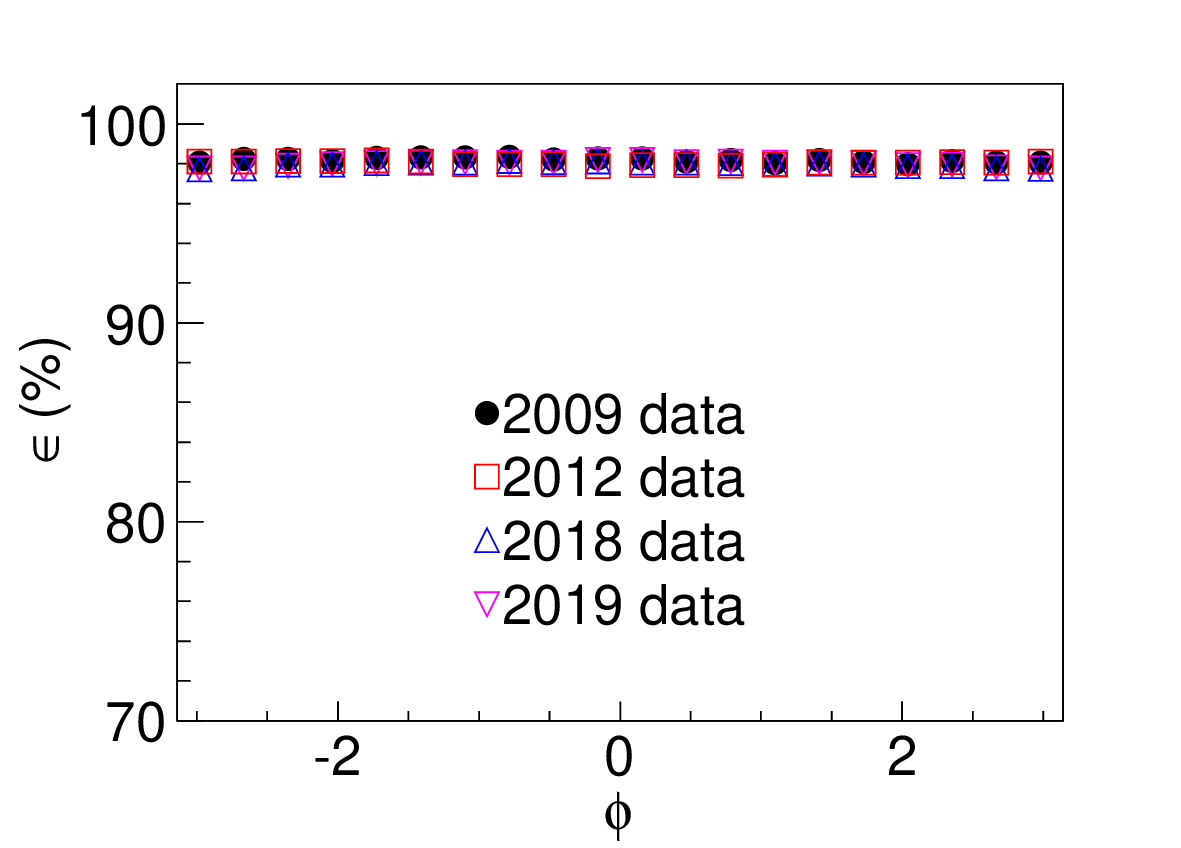}
 \put(-300,50){\bf (a)}\put(-180,50){\bf (b)}\put(-70,50){\bf (c)}\\
  \caption{Pion tracking efficiencies versus $p_t$ (a), $\cos\theta$ (b) and $\phi$ (c) for four data sets.}
 \label{fig:trkeff-1deff-4data-sets}
\end{figure*}

\begin{figure*}[h]
\centering
 \includegraphics[width=0.32\textwidth]{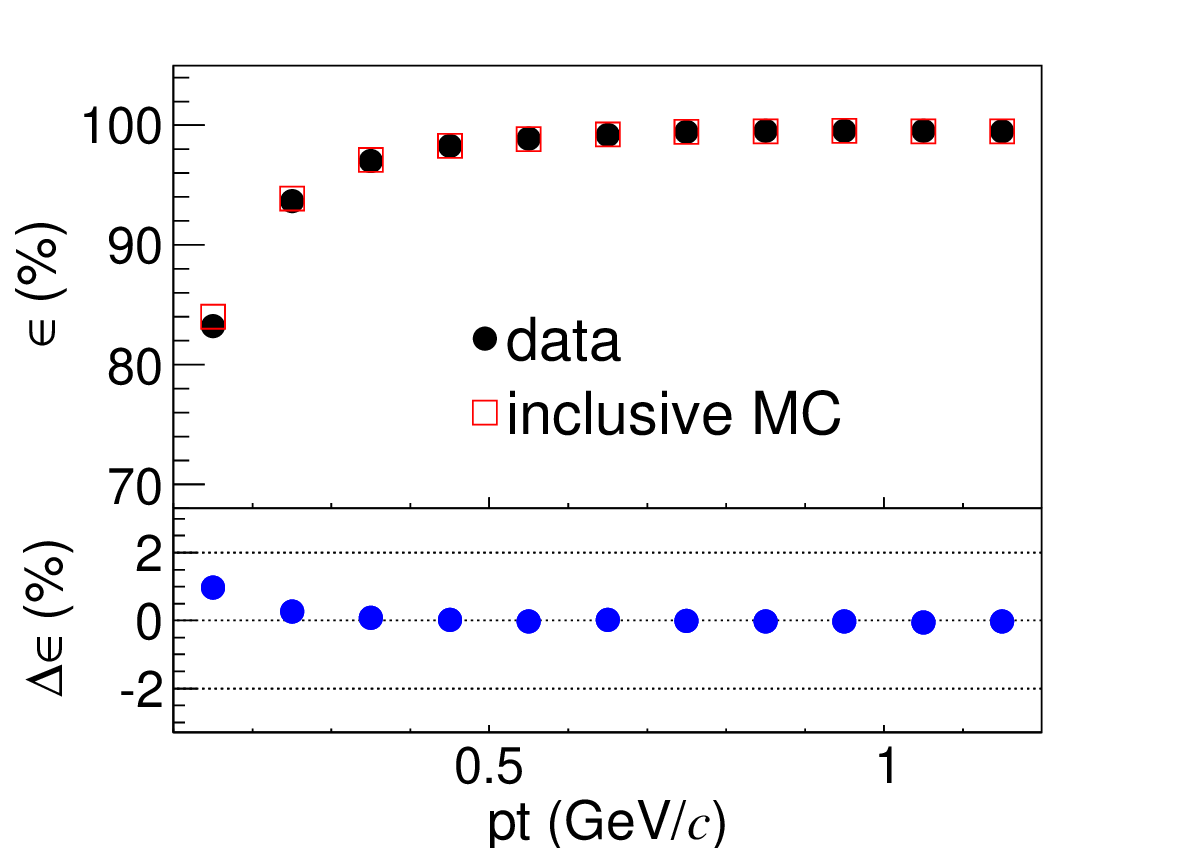}
\includegraphics[width=0.32\textwidth]{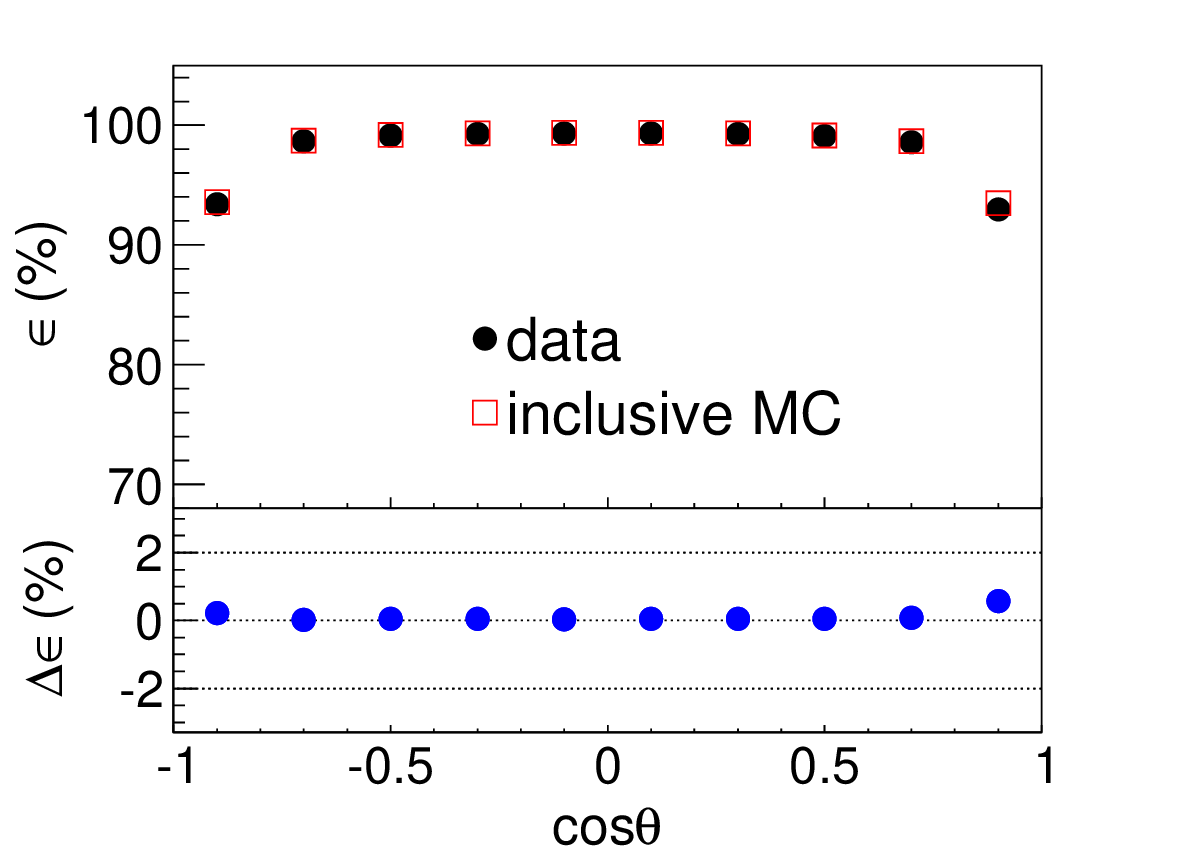}
\includegraphics[width=0.32\textwidth]{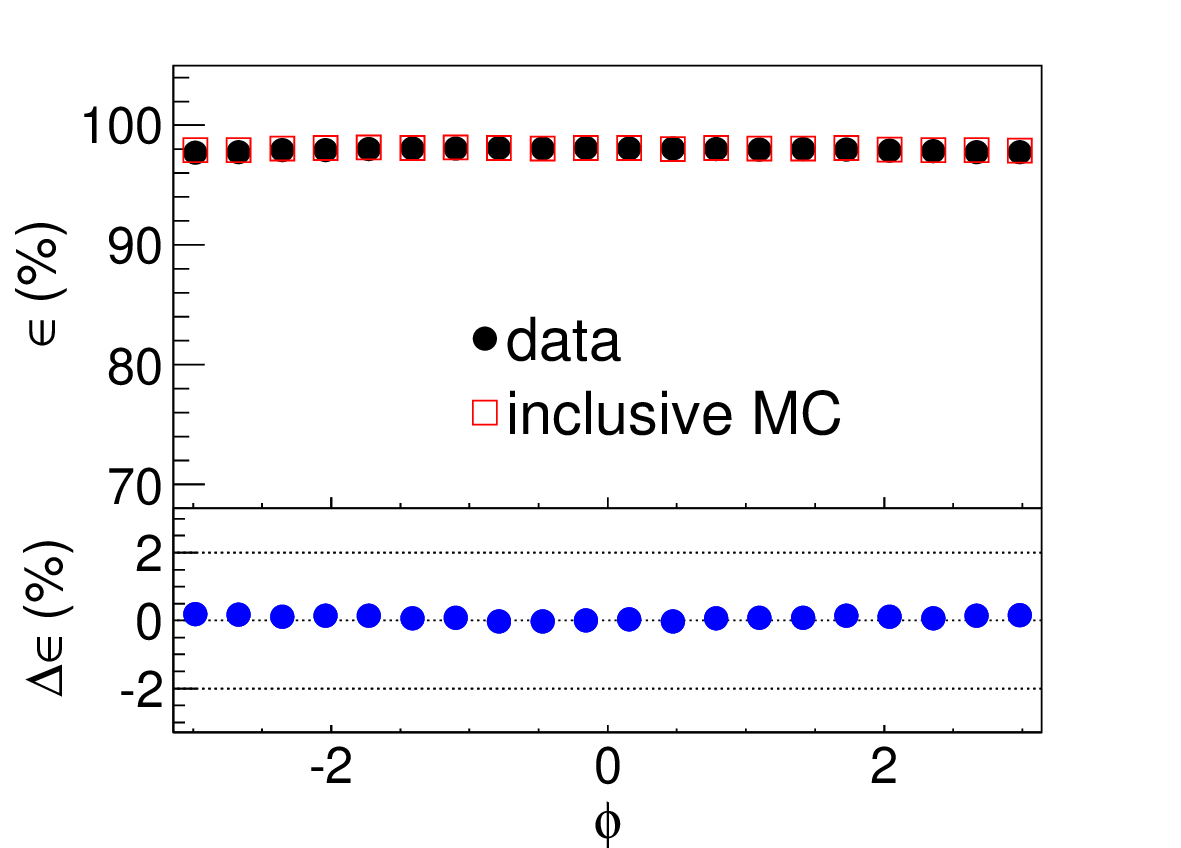}
 \put(-310,60){\bf (a)}\put(-195,60){\bf (b)}\put(-70,60){\bf (c)}
  \caption{Pion tracking efficiencies versus $p_t$ (a), $\cos\theta$ (b) and $\phi$ (c) for 2018 data set. Here $\Delta \epsilon$ is defined as $\frac{\epsilon_{\rm inMC}-\epsilon_{\rm data}}{\epsilon_{\rm inMC}}$ for the embedded bottom plot.}
 \label{fig:trkeff-1deff-datainmc_2018set}
\end{figure*}

The one-dimensional tracking efficiencies versus $p_t$, $\cos\theta$
or $\phi$ for 2009, 2012, 2018 and 2019 data are shown
in Fig.~\ref{fig:trkeff-1deff-4data-sets}. Some difference are visible in the low transverse momentum and 
end cap regions, especially with the higher tracking efficiency of 2009 data set than other data sets, since the MDC accurately measuring  the position and the momentum of the charged particles is suffering from aging issues due to beam related background during the last decades of operation~\cite{dongml2016,dongmy2024}. Pion tracking efffciencies of four data subsets between data and inclusive MC are investigated respectively, and their
systematic uncertainties are also given in Fig.~\ref{fig:trkeff-1deff-datainmc_2018set} (as an
example using 2018 data subset later). The tracking
efficiencies are dependent on $p_t$ and $\cos\theta$ due to different
tracking bendings and hit positions in the MDC, but they are
insensitive to $\phi$.

\begin{figure*}[h]
\centering
 \includegraphics[width=0.32\textwidth]{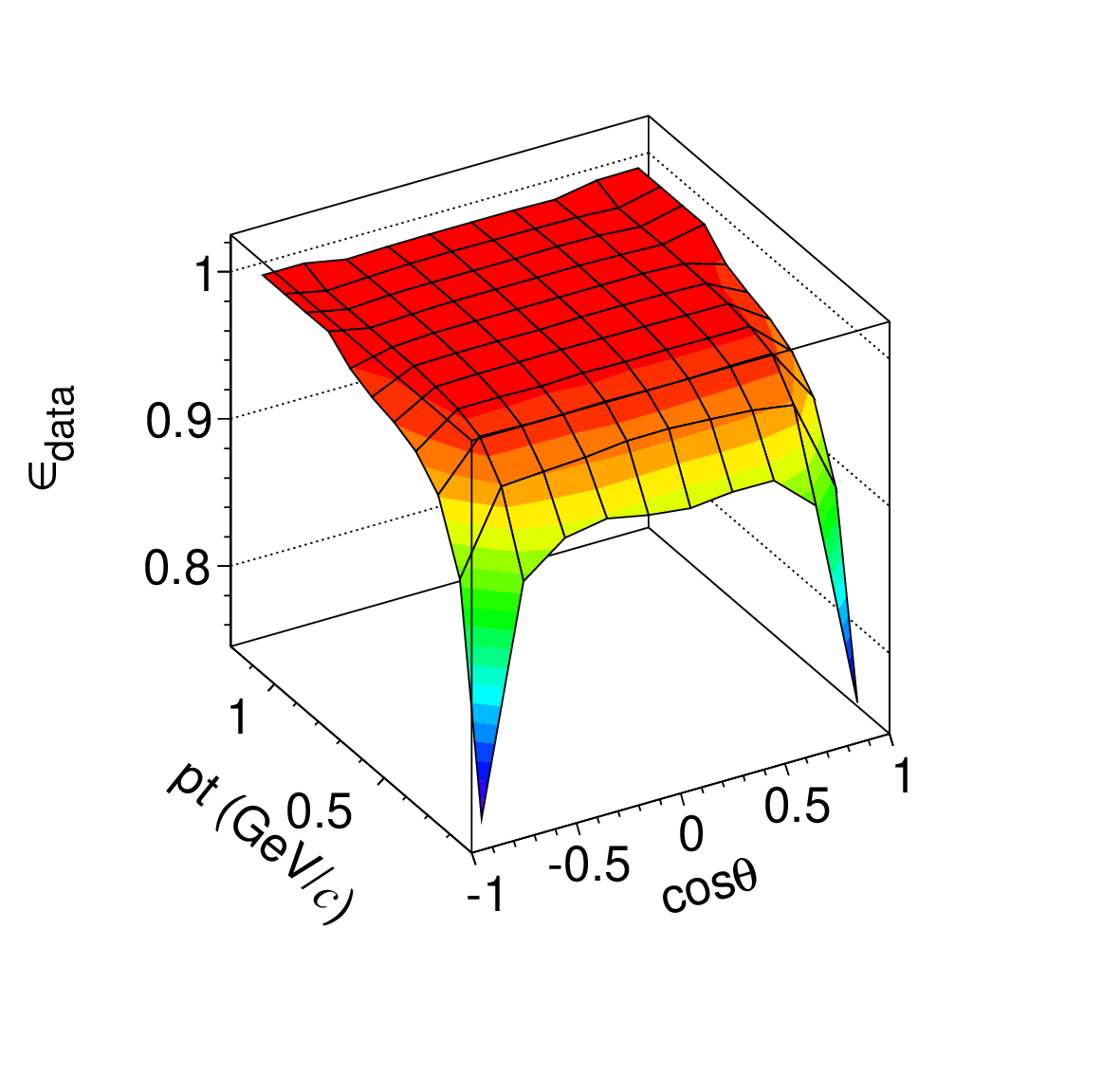}
  \includegraphics[width=0.32\textwidth]{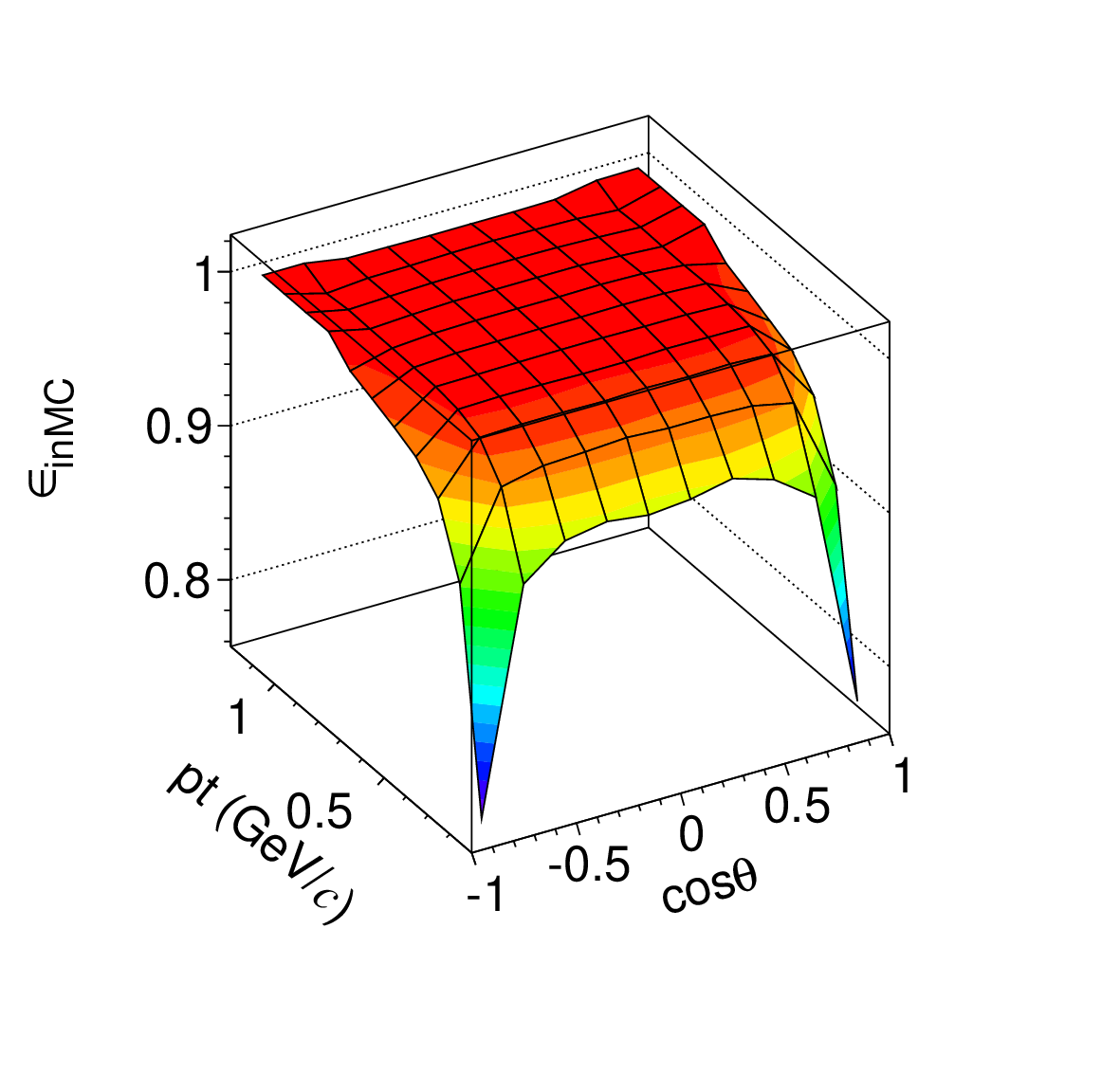}
\includegraphics[width=0.32\textwidth]{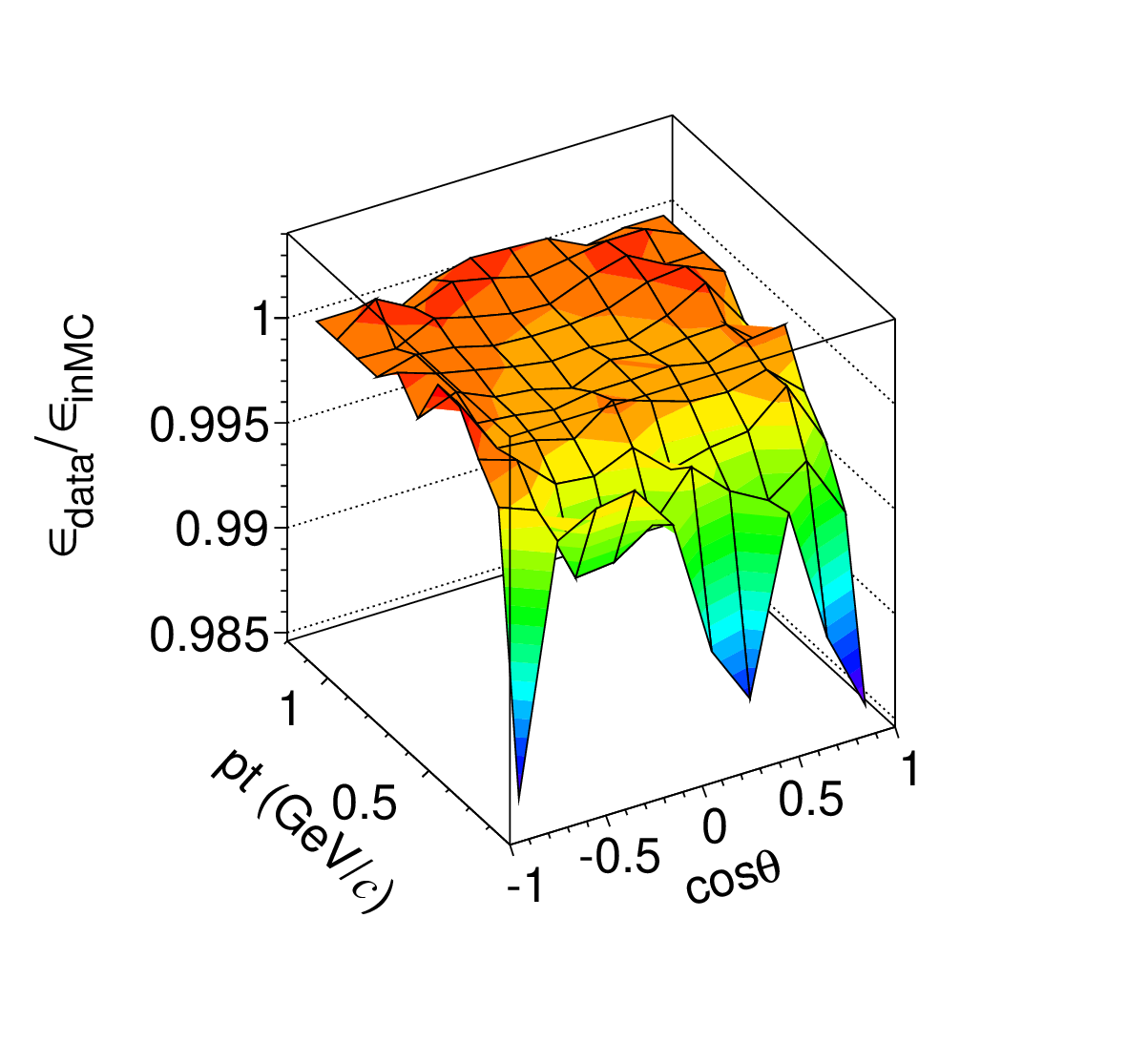}
 \put(-300,35){\bf (a)}\put(-180,35){\bf (b)}\put(-60,35){\bf (c)}
  \caption{ Pion  two-dimensional tracking efficiencies versus
$p_t$ and $\cos\theta$ for (a) data and (b) inclusive MC for 2018 data set. The correction factors are calculated as the efficiency ratios in (c).
  }
 \label{fig:trkeff-2d-data-inmc-2018}
\end{figure*}

Different decay processes exhibit varying data-MC differences in tracking, attributed to distinct distributions in $p_t$ and $\cos\theta$. 
For a given charged particle with fixed transverse momentum and polar angle in the same data set, the tracking efficiency is expected to be the same.
The tracking efficiencies of data and inclusive MC as well as the data-MC differences are parameterized with a function of charge, transverse momentum and polar angle.
With the obtained two-dimensional ($p_t$ versus  $\cos\theta$) tracking efficiency, the systematic uncertainty for a decay process can be obtained after re-weighting them to the corresponding signal MC sample.
As an example, the pion two-dimensional tracking efficiencies of $\pi^{+}$
for data and inclusive MC in  2018 data set are shown in Fig.~\ref{fig:trkeff-2d-data-inmc-2018}(a) and Fig.~\ref{fig:trkeff-2d-data-inmc-2018}(b).
The corresponding correction factors are calculated from the
efficiency ratios of data over inclusive MC, as shown in
Fig.~\ref{fig:trkeff-2d-data-inmc-2018}(c).

The systematic uncertainties of the correction factors on the pion tracking
efficiency mainly come from final event selection and the background.
The systematic uncertainty due to event selection is estimated by the largest change by varying the following selection criteria:
(1) The uncertainty from $\pi^0$ mass window is assigned by varying the
nominal one to  $0.12$ GeV$/c^{2}<M_{\gamma\gamma}<0.145$
GeV$/c^{2}$ or $0.10$ GeV$/c^{2}<M_{\gamma\gamma}<0.16$ GeV$/c^{2}$;
(2) The uncertainty from $\pi^0$ decay
angle is estimated by changing the decay
angle from $\cos\theta<0.95$ to $\cos\theta<0.98$;
(3) The uncertainty from $M^{\rm rec}_{K\pi^0}$
is obtained
with $M^{\rm rec}_{K\pi^0}<0.3$ or 0.4 GeV$/c^{2}$;
(4) The uncertainty from MUC hit layer requirement is obtained by changing the nominal one to be
$N_{\rm first-hit-layer}<2$;
(5) The uncertainty from $E/p$ requirement
is estimated by
varying the nominal requirement to be less than 0.75 or 0.85;
(6) The uncertainty from background is estimated by increasing or reducing the nominal background level by $10\%$.

\begin{figure}[h]
\centering
 \includegraphics[width=0.32\textwidth]{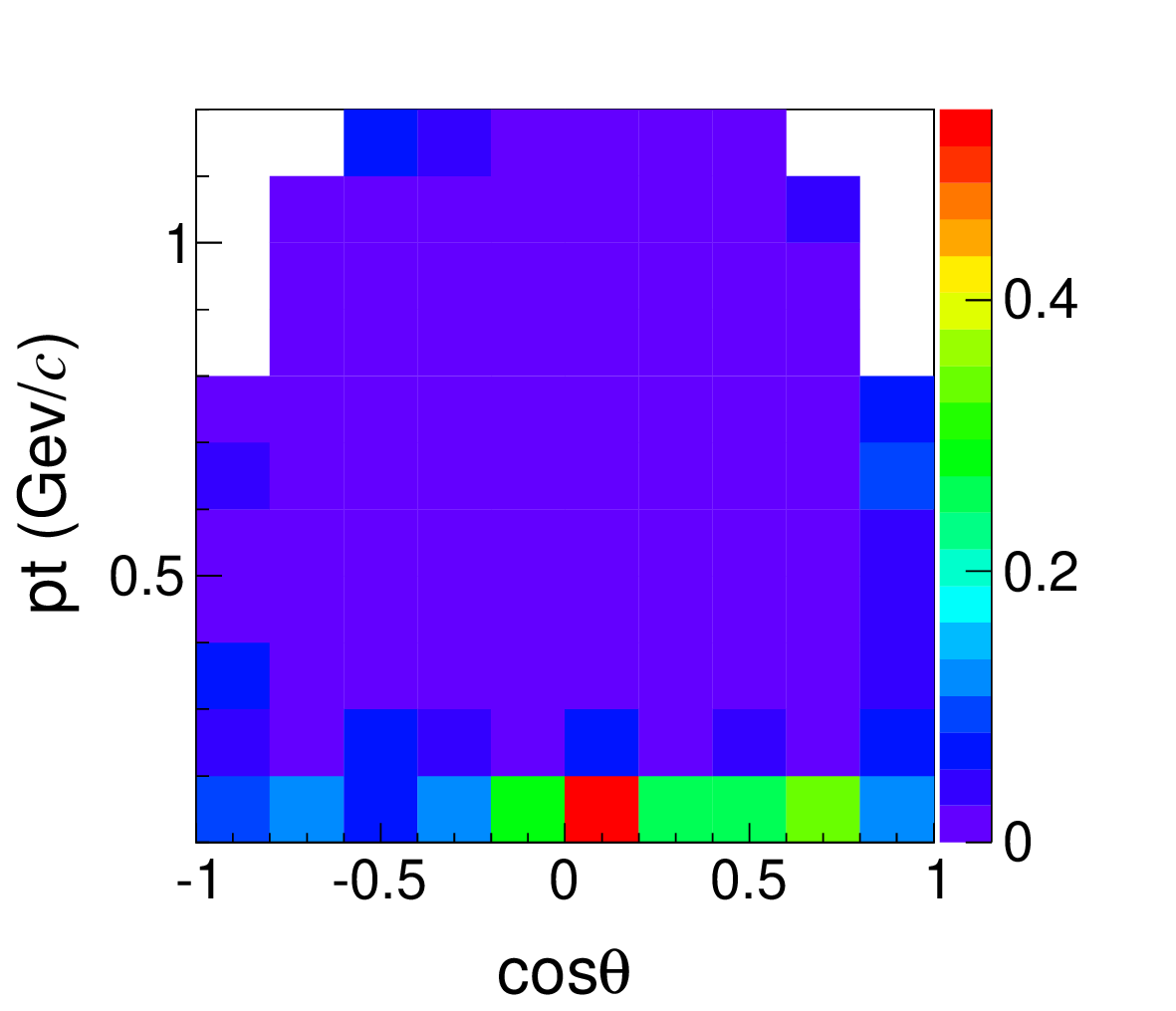}
  \caption{ Pion systematic uncertainty (unit in $\%$) of correction factor on tracking efficiency versus $p_t$ and $\cos\theta$ for 2018 data set.}
 \label{fig:correction-factor-systematicerror-2018}
\end{figure}

The total systematic uncertainty
is given by the quadratic
sum of the individual ones, assuming all sources to be
independent.
 The obtained results are shown in Fig.~\ref{fig:correction-factor-systematicerror-2018}. In most bins, the uncertainties are about $0.1\%$ while they are up to $0.5\%$ in low transverse momentum bins.

\section{Validation of the Tracking Efficiency Correction}
With the above two-dimensional correction factors on tracking efficiency from the control sample,
the corresponding correction factor can be obtained according to the transverse
momentum and polar angle of signal MC sample and weighted on its MC sample, then the differences of efficiencies between data and MC will be estimated.

\begin{figure*}[h]
\centering
 \includegraphics[width=0.32\textwidth]{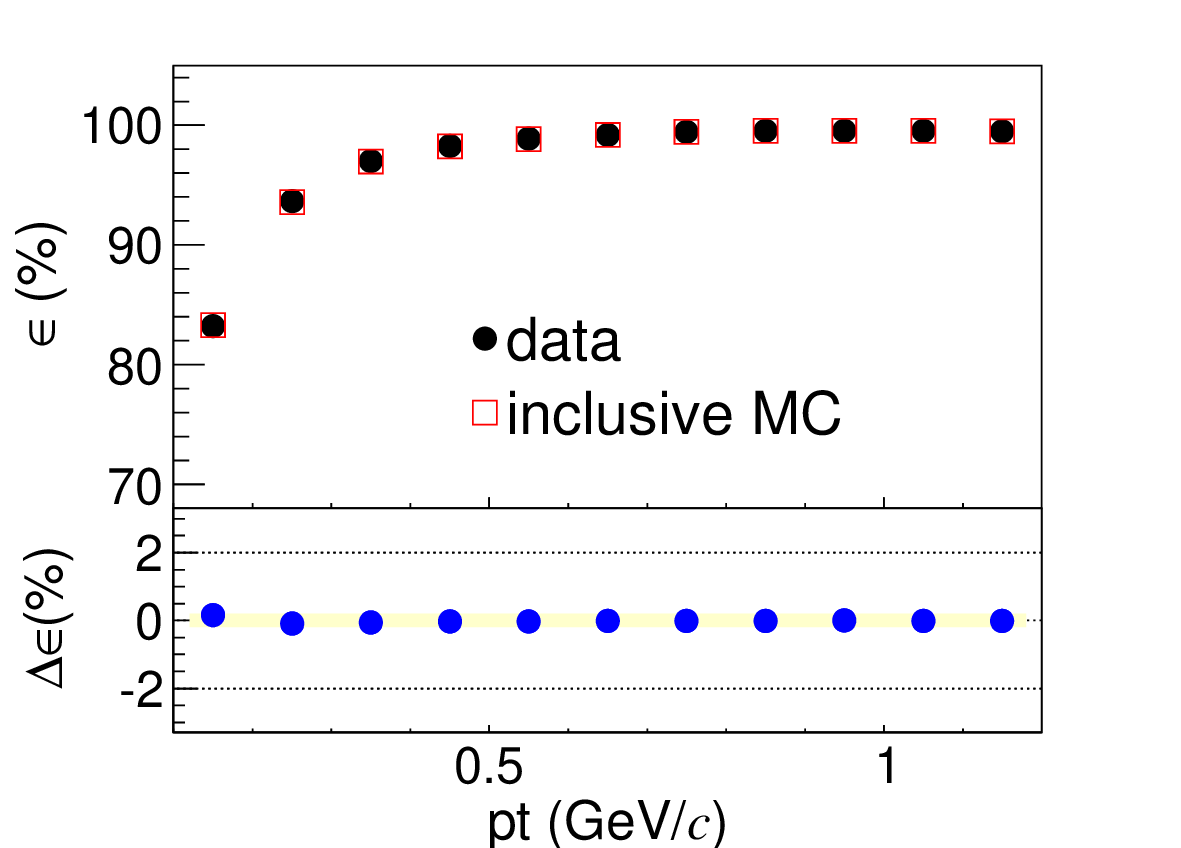}
 \includegraphics[width=0.32\textwidth]{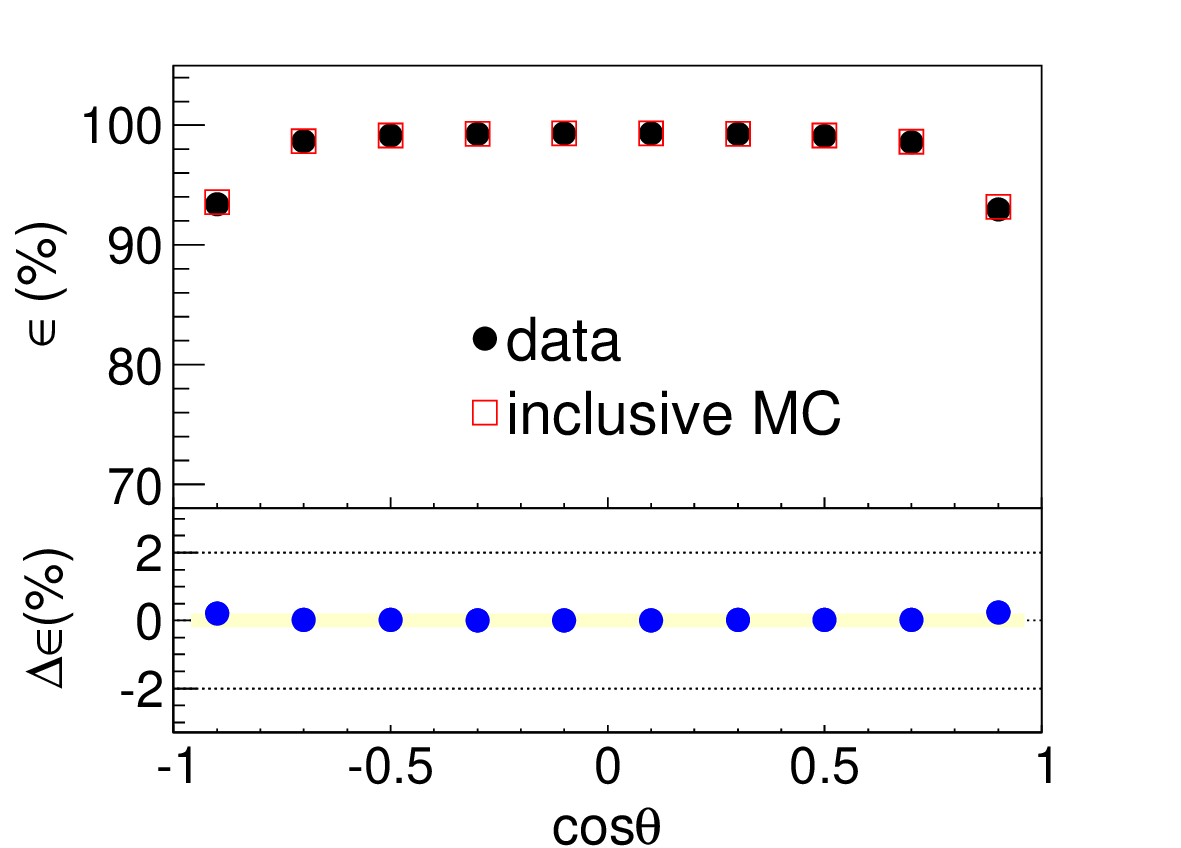}
 \put(-190,60){\bf (a)}\put(-70,60){\bf (b)}\\
 \caption{Tracking efficiencies of pion  sample for 2018 data set after correction.}
 \label{fig:after-2d-correction-factor-pip-tracking-eff}
\end{figure*}

After correcting the tracking efficiency of MC to data for the control sample of $J/\psi \rightarrow \pi^+ \pi^- \pi^0$, the systematic uncertainty of tracking for charged pions
is shown in Fig.~\ref{fig:after-2d-correction-factor-pip-tracking-eff}. The difference is almost zero after correction, thereby validating the method.
If applying these correction parameters on the decay channel of $J/\psi \rightarrow \gamma \eta^{\prime} (\eta^{\prime} \rightarrow \gamma \pi^{+} \pi^{-})$, the systematical uncertainty due to tracking of charged pion can be reduced from $0.4\%$ to $0.2\%$.

\section{Summary}
In summary, using $(10087 \pm 44) \times 10^6$ $J/\psi$ events
collected with the BESIII detector, the tracking efficiency of charged
pions and the systematic uncertainty are
investigated via the decay $J/\psi \rightarrow \pi^{+} \pi^{-} \pi^{0}$. The
corresponding correction factors for data over MC are provided.  The
obtained results are important to reduce the systematic uncertainty
due to tracking in various measurements.

\bmhead{Acknowledgements}
The BESIII Collaboration thanks the staff of BEPCII and the IHEP computing center for their strong support. This work is supported in part by National Key R$\&$D Program of China under Contracts Nos.
2020YFA0406300, 2020YFA0406400, 2023YFA1606000; National Natural Science Foundation of China (NSFC) under Contracts Nos. 11775245, 12075250, 12175259, 12225509, 12275296, 12275297.

\noindent conflict of interest: the authors declare that there is no conflict of interest.


\begin{thebibliography}{99}
\bibitem{Ablikim:2009aa}
  M.~Ablikim {\it et al.} (BESIII Collaboration),
  \href{https://arxiv.org/ftp/arxiv/papers/0911/0911.4960.pdf}{Nucl. Instr.  Meth. Phys. Res.\ Sect.\ A {\bf 614}, 345 (2010).}

\bibitem{Yu:IPAC2016-TUYA01}
   C.~H.~Yu {\it et al.},
  \href{http://jacow.org/ipac2016/papers/tuya01.pdf}{
  Proceedings of IPAC2016, Busan, Korea, 2016.}

\bibitem{PhysRevLett.129.131801}
  M.~Ablikim {\it et al.} (BESIII Collaboration),
  \href{https://arxiv.org/pdf/2204.11058}{Phys. Rev. Lett. {\bf 129}, 131801 (2022).}

\bibitem{zhaozh2024}
  M.~Ablikim {\it et al.} (BESIII Collaboration),
  \href{https://arxiv.org/pdf/2311.12895}{Phys. Rev. D {\bf 109}, 032006 (2024).}
 
 \bibitem{PhysRevD.103.072006}
  M.~Ablikim {\it et al.} (BESIII Collaboration),
  \href{https://arxiv.org/pdf/2012.04257}{Phys. Rev. D {\bf 103}, 072006 (2021).}
 
\bibitem{yanghx} 
M.~Ablikim {\it et al.} (BESIII Collaboration),
\href{https://arxiv.org/pdf/2111.07571.pdf}{Chin. Phys. C {\bf 46}, 074001 (2022).}

\bibitem{bszou2003} 
B. S. Zou and D. V. Bugg,
\href{https://doi.org/10.1140/epja/i2002-10135-4}{Eur. Phys. J {\bf 16}, 537 (2003).}

\bibitem{kkmc2000}
  S. Jadach, B.F.L. Ward and Z. Was,
  \href{https://arxiv.org/pdf/hep-ph/9912214}{Comput. Phys. Commun. {\bf 130}, 260 (2000).}

\bibitem{kkmc2001}
  S. Jadach, B.F.L. Ward and Z. Was,
  \href{https://arxiv.org/pdf/hep-ph/0006359}{Phys. Rev. D {\bf 63}, 113009 (2001).}

\bibitem{2001Lange}
D. J. Lange,
\href{https://inspirehep.net/literature/560129}{Nucl. Instrum. Meth. \textbf{A 462}, 152 (2001).}

\bibitem{2008ping}
R. G. Ping,
\href{https://iopscience.iop.org/article/10.1088/1674-1137/32/8/001}{Chin. Phys. \textbf{C 32}, 599 (2008).}

\bibitem{2022pdg}
S. Navas {\it et al}. (Particle Data Group),
\href{https://journals.aps.org/prd/abstract/10.1103/PhysRevD.110.030001}{Phys. Rev. D \textbf{110}, 030001 (2024).}

\bibitem{2000chen}
J. C. Chen, G. S. Huang, X. R. Qi, D. H. Zhang and Y. S. Zhu,
\href{https://journals.aps.org/prd/abstract/10.1103/PhysRevD.62.034003}{Phys. Rev.  \textbf{D 62}, 034003 (2000).}
\bibitem{lund2014}
 R. L. Yang, R. G. Ping, and H. Chen,
\href{https://iopscience.iop.org/article/10.1088/0256-307X/31/6/061301}{Chin. Phys. Lett. \textbf{ 31},
061301 (2014).}


\bibitem{geant4}
J. Allison {\it et al}. (Geant4 collaboration), 
\href{https://ieeexplore.ieee.org/stamp/stamp.jsp?tp=&arnumber=1610988}{IEEE Trans. Nucl. Sci. \textbf{53} 270 (2006).} 
\bibitem{geant4-2003}
S. Agostinelli {\it et al}. (Geant4 collaboration), 
"Geant4: A simulation toolkit", 
Nucl. Instrum. Meth. A \textbf{506}, 250 (2003)
\href{https://inspirehep.net/files/6c9c0b62bbc8dc0401fca11a5fe5c87c}
{Nucl. Instrum. Meth. \textbf{A 506} 250-303 (2003).}

\bibitem{dongml2016}
M.Y. Dong \textit{et al}.,
\href{https://arxiv.org/pdf/1504.04681}{Chin. Phys. C {\bf 40}, 016001 (2016).
}

\bibitem{dongmy2024}
M.Y. Dong \textit{et al}.,
\href{https://inspirehep.net/literature/2807573}{Nucl. Instrum. Meth. A \textbf{1066}, 169582 (2024).
}


\end{thebibliography}

\end{document}